\documentclass[aps,pra,reprint,superscriptaddress]{revtex4-2}

\usepackage{amsmath,amssymb}
\usepackage{dcolumn}
\usepackage{graphicx}
\usepackage{braket}
\usepackage{float}
\usepackage{microtype}

\usepackage[colorlinks=true,linkcolor=blue,citecolor=blue,urlcolor=blue]{hyperref}

\begin{document}

\title{Higgs and Nambu-Goldstone modes in a spin-1 $XY$ model with long-range interactions}

\author{Daiki Kawasaki} 
\email{daiki.kawasaki@kindai.ac.jp}
\affiliation{Department of Physics, Kindai University, Higashi-Osaka, Osaka 577-8502, Japan}

\author{Ippei Danshita} 
\email{danshita@phys.kindai.ac.jp}
\affiliation{Department of Physics, Kindai University, Higashi-Osaka, Osaka 577-8502, Japan}

\date{\today}

\begin{abstract}
We study theoretically the collective excitations in a spin-1 $XY$ model with a quadratic Zeeman term and a long-range interaction that decays algebraically with the distance.  Using the quantum-field theory based on the finite-temperature Green's function formalism, we analyze properties of the Nambu--Goldstone (NG) and Higgs modes in order to analytically evaluate the damping rate of the Higgs mode in the $XY$-ferromagnetic ordered phase near the quantum phase transition to the disordered phase. When the power of the algebraic decay is 3 as in the case of dipole-dipole interactions in Rydberg-atom systems, we show that at two dimensions the excitation energy of the Higgs mode exhibits a linear dispersion whereas the dispersion of the NG mode becomes proportional to the square root of the momentum.
We find that the damping of the Higgs mode is significantly suppressed by the long-range interaction. We also propose how to excite and probe the Higgs mode in Rydberg-atom experiments.
\end{abstract}

\maketitle

\section{Introduction} \label{sec:Introduction}
The studies of quantum many-body systems with long-range interactions have been recently revitalized by a surge of experimental breakthroughs in state-of-the-art quantum platforms~\cite{Defenu2023}, such as Rydberg atom arrays \cite{Browaeys2020,Morgado2021}, dipolar quantum gases \cite{Lahaye2009,Chomaz2022}, and trapped ions \cite{Blatt2012,Monroe2021}.
Among these platforms, systems of Rydberg atom arrays combine strong dipole-dipole interactions with high controllability and detectability of individual atoms in order to serve as quantum simulators of several quantum many-body models with long-range interactions~\cite{Browaeys2020}.
While previous experiments have mainly focused on quantum simulations of spin-1/2 quantum Ising \cite{Labuhn2016,Bernien2017,Scholl2021} and \textit{XY} models \cite{Chen2025,Chen2023,Sylvain2019}, recent experimental advances have enabled quantum simulations involving three Rydberg states~\cite{Chew2022,Qiao2025}, paving a way toward the quantum simulation of spin-1 models~\cite{Yoshida2024,Moegerle2025}.

This situation allows for exploring a variety of intriguing many-body phenomena in spin-1 models.
Of particular interest among them is the emergence of the Higgs mode, which is absent in homogeneous spin-1/2 systems.
The Higgs amplitude mode is a ubiquitous collective excitation with an energy gap in systems with particle-hole symmetry and spontaneous breaking of a continuous symmetry~\cite{Pekker2015,Shimano2020,Tsuji2024}. It has been experimentally studied in various systems, such as superconducting states of ${\rm NbSe}_2$~\cite{Sooryakumar1980,Sooryakumar1981,Balseiro1980,Littlewood1981,Littlewood1982,Measson2014}, ${\rm Nb}_{1-x}{\rm Ti}_x{\rm N}$~\cite{Matsunaga2013,Matsunaga2014,Sherman2015,Matsunaga2017,Katsumi2024}, ${\rm MgB}_2$\cite{Katsumi2025}, and cuprates~\cite{Katsumi2018, Chu2020}, quantum antiferromagnets consisting of spin dimers ${\rm TlCuCl}_3$~\cite{Ruegg2008,Merchant2014}
 and ${\rm KCuCl}_3$~\cite{Kuroe2012}, charge-density-wave materials ${\rm K}_{0.3}{\rm MoO}_3$~\cite{Demsar1999,Schaefer2014}
 and ${\rm TbTe}_3$~\cite{Yusupov2010,Mertelj2013}, superfluid ${}^{3}{\rm He}$ B-phase~\cite{Avenel1980,Collett2013}, superfluid Bose gases in optical lattices~\cite{Bissbort2011,Endres2012}, supersolid Bose gases~\cite{Leonard2017,Hertkorn2019}, and superfluid Fermi gases~\cite{Behrle2018,Dyke2024,Kell2024,Cabrera2025}.
In an intuitive picture, the Higgs mode corresponds to a fluctuation of the amplitude of the order parameter.
This mode in condensed-matter and cold-atom systems is named after the Higgs boson in particle physics~\cite{Higgs1964}, because they are analogous within the quantum-field theoretical description in continuum.

In two-dimensional systems with short-range interactions, strong effects of quantum and thermal fluctuations typically lead to severe damping of the Higgs mode~\cite{Altman2002,Podolsky2011,Pollet2012,Podolsky2012,Rancon2014}, making its experimental detection rather difficult. 
Experiments with two-dimensional ultracold Bose gases in optical lattices, which is quantitatively described by the Bose-Hubbard model with the nearest-neighbor hopping, have indeed attempted to detect the Higgs mode in the superfluid phase near the quantum phase transition to the Mott insulator~\cite{Endres2012}.
They have observed the response of the system to a temporal modulation of the optical-lattice amplitude. While they have successfully measured the gap energy of the Higgs mode, the response as a function of the modulation frequency has not exhibited a sharp peak, which is a smoking gun of long-lived elementary excitation, but a broad continuum. On the other hand, in systems with the long-range interactions decaying with the distance $r$ as $\propto r^{-\alpha}$, effects of the fluctuations are in general weaker than short-range interacting systems as manifested in the fact that long-range ordered phases breaking spontaneously a continuous symmetry can be a thermal equilibrium state even at finite temperatures and two dimensions when $\alpha<4$~\cite{Bruno2001,Sbierski2024}. This raises an interesting question of whether the Higgs mode can be long-lived in the long-range interacting systems at two dimensions.

To address this question, we investigate theoretically properties of the collective excitations in the $XY$-ferromagnetic ordered phase of a spin-1 \textit{XY} model with the long-range interactions and a quadratic Zeeman term, which approximately describes an atom array with effective three Rydberg states~\cite{Chew2022,Yoshida2024}.
Using the mean-field and quantum-field-theoretical approaches~\cite{Nagao2016,Nagao2016b}, we analytically calculate the dispersion relations of the Higgs and Nambu--Goldstone (NG) modes, and the Beliaev damping rate of the Higgs mode as functions of the spatial dimension $d$ and the power of the long-range interaction $\alpha$.
We find that when $d=2$ and $\alpha=3$, corresponding to the system of the Rydberg-atom array, the long-range interaction strongly suppresses the damping of the Higgs mode so that the Higgs mode can be long-lived even at finite temperatures. We also propose a way to create and probe the Higgs mode in Rydberg-atom experiments. 

The remainder of the paper is organized as follows: In Sec.~\ref{sec:Model}, we explain the spin-1 \textit{XY} model with a quadratic Zeeman term and the long-range interaction and review the ground state phase diagram of the model in the case of the short-range interaction.
In Sec.~\ref{sec:Methods}, we present our theoretical methods describing  properties of the ground state and collective excitations.
In Sec.~\ref{sec:Results}, we show our results focusing on the dispersion relations of the NG and Higgs modes, and the Beliaev damping of the Higgs mode.
In Sec.~\ref{sec:Conclusion}, we conclude the paper with summary and outlook.

Throughout this paper, we set $\hbar = d_{\mathrm{lat}} = k_{\mathrm{B}} = 1$, where $\hbar$ is the reduced Planck constant, $d_{\mathrm{lat}}$ the lattice spacing, and $k_{\mathrm{B}}$ the Boltzmann constant. 

\section{Model} \label{sec:Model}
We consider a system of Rydberg atoms arranged in a square lattice form by an optical-tweezer array. We focus on a situation in which the system is well described by  
the $S=1$ ferromagnetic \textit{XY} model with linear and quadratic Zeeman terms as well as long-range spin-exchange interactions~\cite{Yoshida2024},
\begin{align}
  \hat{H} =& - \sum_{j < l} J_{j,l}(\alpha=3) \left(\hat{S}^{+}_{j} \hat{S}^{-}_{l} + \mathrm{H.c.} \right)
  \nonumber \\
  &+ \Delta \sum_{j} (\hat{S}^{z}_{j})^{2} - p \sum_{j} \hat{S}^{z}_{j}, \label{eq:Hamiltonian}
\end{align}
where $\hat{S}^{\pm}_{j}$ and $\hat{S}^{z}_{j}$ are the spin-1 operators at site $j$, $J_{j,l}(\alpha) = J/|\mathbf{r}_{j} - \mathbf{r}_{l}|^{\alpha}$ the long-range spin-exchange interaction between sites $j$ and $l$, $J(>0)$ the nearest-neighbor spin-exchange interaction, $\mathbf{r}_{j}$ the position of site $j$, $\Delta$ the quadratic Zeeman coefficient, 
and $p$ the linear Zeeman coefficient.
Such a situation can be realized by utilizing three different Rydberg states which can be exchanged with one another via three-state F\"orster coupling resulting from the dipole-dipole interactions~\cite{Ravets2014,Chew2022}. For example, in experiments of Ref.~\cite{Chew2022}, the three states $|41F\rangle$, $|43D\rangle$, and $|45P\rangle$ of the $^{87}$Rb atom, whose single-atom eigenenergies are $E_{41F}$, $E_{43D}$, and $E_{45P}$, have been utilized and a coherent Rabi oscillation between $|43D,43D\rangle$ and $|41F,45P\rangle+|45P,41F\rangle$ has been observed in a two-atom system.
In terms of the spin-1 model, the Rydberg states $|41F\rangle$, $|43D\rangle$, and $|45P\rangle$ correspond to the local spin states $|S^z=1\rangle$, $|0\rangle$, and $|-1\rangle$, the F\"orster coupling to the $XY$ spin-exchange interaction, and the detuning from the F\"orster resonance to the quadratic Zeeman term as $\Delta=\frac{1}{2}\left(E_{41F}+E_{45P}-2E_{43D}\right)$.

In actual experiments~\cite{Chew2022}, the state dependence of the strength of the dipole-dipole interactions leads to deviation from the simple $XY$ exchange interactions of the Hamiltonian (\ref{eq:Hamiltonian}) such that some additional exchange terms exist~\cite{Yoshida2024}. Nevertheless, since this deviation is relatively small, we neglect the additional terms in the following analyses. While the power $\alpha$ of the power-law decaying interaction is fixed to be $\alpha =3$ in the Rydberg-atom system, we vary $\alpha$ in the region $d<\alpha <d+2$ for theoretical interest, where $d$ is the spatial dimension. While the condition $d> 2$ is incompatible with the spatially isotropic dipole-dipole interactions, we also regard $d$ as a variable  under the assumption that the lattice structure is hypercubic.
The quadratic Zeeman coefficient $\Delta$ can be controlled by exposing atoms to a global laser that induces the light shift of one of the three states~\cite{Ravets2014}. We assume that $\Delta >0$.

\begin{figure}[t]
  \centering
  \includegraphics[width=\columnwidth]{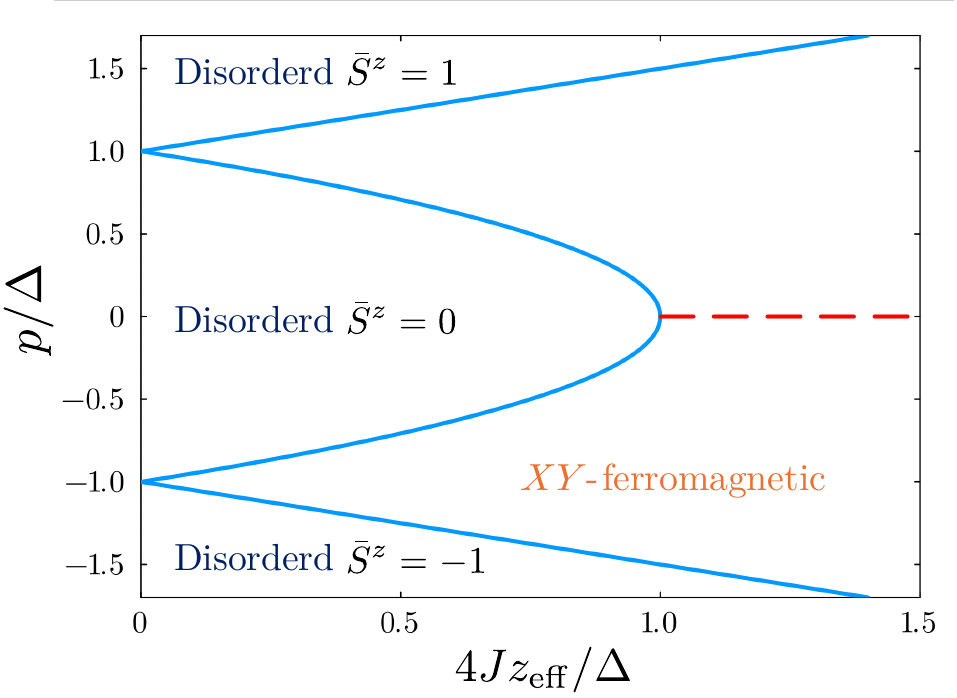}
  \caption{Ground-state phase diagram of the spin-1 $XY$ model of Eq.~(\ref{eq:Hamiltonian}) obtained within the mean-field approximation for $\alpha>d$. The blue solid line represents the phase boundary between the $XY$-ferromagnetic ordered phase and the disordered phases. In the analogy to the Bose-Hubbard model at large filling~\cite{Altman2002,Nagao2016,Nagao2016b}, $\bar{S}^{z}$ corresponds to the deviation of the local occupation from a reference commensurate filling $n_{0}$: the disordered region with $\bar{S}^{z}=0$ is analogous to the Mott-insulator phase with filling $n_{0}$, whereas the regions with $\bar{S}^{z}=\pm1$ correspond to the neighboring Mott lobes with fillings $n_{0}\pm1$. The red dashed line indicates the line of $\frac{p}{\Delta} = 0$ in the ordered phase region.}
  \label{fig:phasediagram}
\end{figure}

When the spin-exchange interaction includes only the nearest-neighbor one, i.e., $\alpha\rightarrow \infty$, the previous theoretical studies have mapped out the ground-state phase diagram of the model of Eq.~(\ref{eq:Hamiltonian}), which is equivalent to the Bose-Hubbard model in the limit of the large filling factor~\cite{Altman2002,Nagao2016,Nagao2016b}.
Figure \ref{fig:phasediagram} reviews the mean-field phase diagram in the $(\frac{4Jz_{\rm eff}}{\Delta},\frac{p}{\Delta})$ plane obtained in Refs.~\cite{Nagao2016b}, where $z_{\rm eff}$ for this case represents the coordination number. 
When $\frac{Jz_{\rm eff}}{\Delta} \gg 1$ or the mean value of $S^z$ per site, which we call $\bar{S}^z$, is noninteger, the ground state forms an $XY$-ferromagnetic ordered phase, 
in which the $XY$ component of all the spins are aligned in a particular direction. 
This phase spontaneously breaks the continuous U(1) symmetry of the Hamiltonian.
When $\frac{Jz_{\rm eff}}{\Delta}$ is decreased along $\frac{p}{\Delta}=0$, where $\bar{S}^z=0$, a continuous quantum phase transition to the disordered phase occurs at a certain critical point. The phase boundary $\left(\frac{p}{\Delta}\right)_{\rm c}$ as a function of $\frac{Jz_{\rm eff}}{\Delta}$ reads $\pm\sqrt{1-\frac{4Jz_{\rm eff}}{\Delta}}$. Introducing a dimensionless parameter $u = \frac{\Delta}{4 J z_{\mathrm{eff}}}$, the critical point $u_{\rm c}$ along $\frac{p}{\Delta}=0$ is given by $u_{\rm c}=1$  within the mean-field approximation. In this sense, $u$ measures the distance from the critical point.
In the mapping to the Bose-Hubbard model at large filling~\cite{Altman2002,Nagao2016,Nagao2016b}, $\bar{S}^{z}$ corresponds to the deviation of the local occupation from a reference commensurate filling $n_{0}$. Accordingly, the disordered phase with $\bar{S}^{z}=0$ is analogous to the Mott-insulator phase with filling $n_{0}$, whereas the disordered phases with $\bar{S}^{z}=\pm1$ are analogous to the neighboring Mott-insulator phases with fillings $n_{0}\pm1$. In this correspondence, the local states with $S^{z}=+1$ and $S^{z}=-1$ represent particle and hole excitations relative to $n_{0}$, respectively.
Notice that more precise values of $u_{\rm c}$ for $d=1$, 2, and 3 have been also computed by means of more sophisticated numerical methods~\cite{Danshita2011,Teichmann2009,Teichmann2009b}. When $\bar{S}^z=\pm1$, the ground state is trivially a disordered phase and the boundary to the $XY$-ferromagnetic phase is given by $\left(\frac{p}{\Delta}\right)_{\rm c}=\pm\left(1+\frac{2Jz_{\rm eff}}{\Delta}\right)$. In the next section, we see that in the case of the long-range interactions ($d<\alpha$), the ground-state phase diagram is identical to the case of the nearest-neighbor one with the replacement of $z_{\rm eff}$ by that for the long-range interaction. 

The previous theoretical studies on the nearest-neighbor case have also shown that the system in the $XY$-ferromagnetic ordered phase for $\bar{S}^z=0$ possesses the Higgs and Nambu--Goldstone (NG) modes, which respectively correspond to fluctuations of the amplitude and phase of the order parameter, as low-energy elementary excitations~\cite{Altman2002,Nagao2016,Nagao2016b}. The presence of the Higgs amplitude mode can be attributed to the particle-hole symmetry at $\frac{p}{\Delta}=0$, i.e., the symmetry with respect to the replacement of $\hat{S}^z_j$ by $-\hat{S}^z_j$. The lifetime of the Higgs mode has been also evaluated to elucidate that the Higgs mode is severely damped at $d=2$ both for zero and finite temperatures due to strong effects of quantum and/or thermal fluctuations~\cite{Altman2002,Podolsky2011,Pollet2012,Podolsky2012,Rancon2014}. In the following sections, we show that the long-range interaction significantly suppresses the effects of the fluctuations so that the Higgs mode can be long-lived even at $d=2$.  

\section{Methods} \label{sec:Methods}
In this section, we present theoretical methods for analyzing low-energy excitations in the $XY$-ferromagnetic phase of the model (\ref{eq:Hamiltonian}) near the transition to the disordered phase, namely, the NG and Higgs modes. Specifically, 
we apply the mean-field and quantum-field-theoretical approaches, which have been used for the case of the nearest-neighbor interaction in Ref.~\cite{Nagao2016,Nagao2016b}, to our system with the long-range interaction.

\subsection{Mean-field approximation} \label{subsec:s2a}
We approximate the quantum many-body state of the system as the following variational state that takes a form of the simple product state~\cite{Altman2002},
\begin{align}
  | \Omega \rangle = &\bigotimes_j \left\{  \cos\left(\frac{\theta_j}{2}\right) | 0 \rangle_{j}  + e^{i \eta_j} \sin\left(\frac{\theta_j}{2}\right) \right. \nonumber \\
  &\left. \times \left[ e^{i \varphi_j} \sin\left(\frac{\chi_j}{2}\right) |1 \rangle_{j} + e^{- i \varphi_j} \cos\left(\frac{\chi_j}{2}\right) |-1 \rangle_{j} \right] \right\} , \label{eq:Var-WF}
\end{align}
where $\theta_j \in [0, \pi]$, $\eta_j \in [-\pi/2, \pi/2]$, $\varphi_j \in [0, 2\pi]$, and $\chi_j \in [0, \pi]$ are the variational parameters specifying the state. While the variational parameters in general depends on $j$, those for the $XY$-ferromagnetic ordered state and the disordered states, which are candidates for the ground state, are homogeneous so we henceforth set $(\theta_j,\eta_j,\varphi_j,\chi_j)=(\theta,\eta,\varphi,\chi)$. The variational parameters for the ground states are determined in a way such that the mean-field energy per site 
\begin{eqnarray}
&&\!\!\!\!\!\!\!\!\!\! E^{\mathrm{MF}} =\frac{1}{M} \langle \Omega | \hat{H} | \Omega \rangle
= \left( \Delta + p \cos \chi \right) \sin^{2} \left( \frac{\theta}{2} \right) 
 \nonumber \\
 &&\,\,\,\,\,\,\,\,\,\,\,\,\,\,\,\,\,\,\,\,\,\,\,\,\,\,\,\,\,\,\,\,\,\,\,\,\,\,\,\,\,\,\,\,- \frac{J z_{\mathrm{eff}}}{2} \sin^{2} \theta \left( 1 + \sin \chi \cos (2 \eta) \right), 
 \label{eq:MeanE}
\end{eqnarray}
is minimized, where $M$ is the total number of sites. In the case of the long-range interaction, the effective coordination number $z_{\rm eff}$ constitutes the summation of all the interaction terms as
\begin{align}
  z_{\mathrm{eff}} = \sum_{l\neq0}\frac{J_{0,l}}{J}, \label{eq:zeff}
\end{align}
where ${\bf r}_l=0$ at $l=0$. For the specific case of $\alpha = 3$ and $d = 2$, which is the most relevant to Rydberg-atom experiments, its numerical value is $9.032$ \cite{Danshita2010}.

The variational state can represent both of the $XY$-ferromagnetic ordered phase and the disordered phases. To see this, we explicitly express the $XY$-ferromagnetic order parameter $\Psi = \langle \Omega | \hat{S}^-_j | \Omega \rangle$ and the mean value of $\hat{S}_j^z$ per site as
\begin{align}
\Psi=\frac{1}{\sqrt{2}} & e^{i\varphi}\sin\theta \left[ e^{i\eta}\sin\left(\frac{\chi}{2}\right) + e^{-i\eta}\cos\left(\frac{\chi}{2}\right) \right],
\label{eq:OP}
\\
&\bar{S}^z=\langle \Omega | \hat{S}^z_j | \Omega \rangle=-\cos\chi\sin^2\left(\frac{\theta}{2}\right).
\label{eq:Sz}
\end{align}
Equations (\ref{eq:OP}) and (\ref{eq:Sz}) imply that the disordered states, where $\Psi=0$, with $\bar{S}^z=1$ and $-1$ correspond to the states with $(\theta,\chi)=(\pi,\pi)$ and $(\pi,0)$ while the other parameters are not determined. That with $\bar{S}^z=0$ corresponds to $\theta=0$. The variational state with $\theta\neq 0$ represents the $XY$-ferromagnetic phase. In this phase, one can easily see from Eq.~(\ref{eq:MeanE}) that $\eta=0$ and $\varphi$ can be chosen arbitrarily. Without loss of generality, we choose $\varphi=0$, which means that the U(1) symmetry of the model (\ref{eq:Hamiltonian}) is spontaneously broken in this phase. Moreover, when $\frac{p}{\Delta}=0$, $\chi=\frac{\pi}{2}$ is satisfied. 

From the Ginzburg-Landau expansion of $E^{\mathrm{MF}}$ of the ground state with respect to the order parameter $\Psi $, one can determine the boundaries between the $XY$-ferromagnetic ordered phase to the disordered phases. The phase boundaries are identical to the case of the nearest-neighbor interaction, which was discussed in the previous section, by extending the coordination number to the effective one $z_{\rm eff}$. Recall that the ground-state phase diagram is shown in Fig.~\ref{fig:phasediagram}.

\subsection{Schwinger-boson representation of the spin-1 operators} \label{subsec:s2b}
The Hamiltonian of Eq.~(\ref{eq:Hamiltonian}) can be represented using three Schwinger bosons \cite{Altman2002,Nagao2016,Huber2007,Wang2021}:
\begin{align*}
  | \mu \rangle_{j} = \hat{t}^{\dagger}_{\mu, j} | \mathrm{vac} \rangle_{j} ~~~\mathrm{for}~ \mu = -1, 0, 1,
\end{align*}
where $\ket{\mathrm{vac}}$ is the vacuum of the new bosons. The commutation relations are $[\hat{t}_{\mu, j},\hat{t}^{\dagger}_{\mu',j'}]=\delta_{\mu, \mu'}\delta_{j,j'}$ and $[\hat{t}_{\mu, j},\hat{t}_{\mu', j'}]=[\hat{t}^{\dagger}_{\mu, j},\hat{t}^{\dagger}_{\mu', j'}]=0$.
In order to eliminate the unphysical states such as $\hat{t}^{\dagger}_{1, j} \hat{t}^{\dagger}_{0, j} | \mathrm{vac} \rangle$, we assume that these operators obey a constraint,
\begin{align}
  \sum_{\mu =-1}^{1} \hat{t}^{\dagger}_{\mu, j} \hat{t}_{\mu, j} = 1, \label{LC1}
\end{align}
where $1$ on the right-hand side is the identity operator.

The spin operators are represented as
\begin{align*}
  \hat{S}^{+}_{j} &= \sqrt{2} (\hat{t}^{\dagger}_{1,j} \hat{t}_{0,j} + \hat{t}^{\dagger}_{0,j} \hat{t}_{-1,j}), \\
   \hat{S}^{z}_{j} &= \hat{t}^{\dagger}_{1,j} \hat{t}_{1,j} - \hat{t}^{\dagger}_{-1,j} \hat{t}_{-1,j}, \\
  (\hat{S}^{z}_{j})^2 &= \hat{t}^{\dagger}_{1,j} \hat{t}_{1,j} + \hat{t}^{\dagger}_{-1,j} \hat{t}_{-1,j}.
\end{align*}
One can check that with this representation they satisfy the SU(2) commutation relations $[\hat{S}^{+}_{j}, \hat{S}^{-}_{l}] = 2\hat{S}^{z}_{j} \delta_{j,l}$ and $[\hat{S}^{z}_{j}, \hat{S}^{\pm}_{l}] = \pm \hat{S}^{\pm}_{j} \delta_{j,l}$.

Let us introduce a canonical transformation in the following manner~\cite{Altman2002}:
\begin{align}
  \hat{b}^{\dagger}_{0, j} &= c_{1} \hat{t}^{\dagger}_{0, j} + \frac{1}{\sqrt{2}} s_{1} \left[\hat{t}^{\dagger}_{1, j} + \hat{t}^{\dagger}_{-1, j} \right], \nonumber \\ 
  \hat{b}^{\dagger}_{\mathrm{A}, j} &= s_{1} \hat{t}^{\dagger}_{0, j} - \frac{1}{\sqrt{2}} c_{1} \left[\hat{t}^{\dagger}_{1, j} + \hat{t}^{\dagger}_{-1, j} \right], \label{CnT} \\ 
  \hat{b}^{\dagger}_{\Phi, j} &= \frac{1}{\sqrt{2}} \left[ \hat{t}^{\dagger}_{1, j} - \hat{t}^{\dagger}_{-1, j} \right], \nonumber
\end{align}
where the coefficients are $s_{1} = \sin\left(\frac{\theta_{\mathrm{gs}}}{2}\right)$ and $c_{1} = \cos\left(\frac{\theta_{\mathrm{gs}}}{2}\right)$. Moreover,
$\theta_{\mathrm{gs}}$ denotes the value of the variational parameter $\theta$ for the ground state, $\hat{b}^{\dagger}_{\mathrm{0}, j}$ the creation operator of the mean-field ground state, and
$\hat{b}^{\dagger}_{\mathrm{A}, j}$ ($\hat{b}^{\dagger}_{\Phi, j}$) the creation operator for the amplitude (phase) fluctuation of the order parameter.
These new operators fulfill the same bosonic commutation relations as $\hat{t}_{\mu, j}$.
In addition, the transformation retains the constraint \eqref{LC1} so that
\begin{align}
  \sum_{m \in \{0, \mathrm{A}, \Phi\}} \hat{b}^{\dagger}_{m, j} \hat{b}_{m, j} = 1. \label{LC2}
\end{align}

Substituting the canonical transformation \eqref{CnT} into the Hamiltonian of Eq.~\eqref{eq:Hamiltonian}, one can express the Hamiltonian in terms of $\hat{b}_{m,j}$. 

\subsection{Holstein--Primakoff expansion} \label{s2d}
In this section, assuming that quantum and thermal fluctuations from the $XY$-ferromagnetic ground state are weak, we expand the Hamiltonian with respect to $\hat{b}_{{\rm A},j}$ and $\hat{b}_{\Phi,j}$ in order to understand the low-energy physics of the system in the framework of the quasiparticle picture, in which the Higgs and NG modes play crucial roles.
Specifically, we use the Holstein--Primakoff expansion~\cite{Holstein1940}:
\begin{align}
  \hat{b}^{\dagger}_{m, j} \hat{b}_{0, j} &= \hat{b}^{\dagger}_{m, j} \sqrt{1 - \hat{b}^{\dagger}_{\mathrm{A}, j} \hat{b}_{\mathrm{A}, j} - \hat{b}^{\dagger}_{\Phi, j} \hat{b}_{\Phi, j}} \label{HPex} \\
&\simeq \hat{b}^{\dagger}_{m, j} - \frac{1}{2}\hat{b}^{\dagger}_{m, j}\hat{b}^{\dagger}_{\mathrm{A}, j} \hat{b}_{\mathrm{A}, j} - \frac{1}{2} \hat{b}^{\dagger}_{m, j}\hat{b}^{\dagger}_{\Phi, j} \hat{b}_{\Phi, j} + \cdots. \nonumber
\end{align}
Eliminating $\hat{b}^{\dagger}_{0, j}$ and $\hat{b}_{0, j}$ in the Hamiltonian by using the constraint \eqref{LC2}, and substituting the Holstein--Primakoff expansion \eqref{HPex} into the Hamiltonian, we obtain the following series:
\begin{align}
  \hat{H} \simeq \hat{H}^{(0)} + \hat{H}^{(1)} + \hat{H}^{(2)} + \hat{H}^{(3)} + \cdots, \label{HPHam}
\end{align}
where $\hat{H}^{(i)}$ (for $i = 0, 1, 2, 3, \ldots$) represents the $i$th order perturbation term with respect to $\hat{b}_{m,j}$.
The expansion \eqref{HPHam} is stopped at $i = 3$.
We keep cubic terms in the expansion to capture the Beliaev damping process, where a Higgs mode decays into two NG modes.
In the case of finite temperatures, the above truncation is valid only at $d \geq 2$, where the long-range $XY$-ferromagnetic order is present in the thermal equilibrium state. Notice that when the interaction is longranged as $d<\alpha<d+2$, the long-range order holds even at $d=2$, i.e., the celebrated Mermin--Wagner theorem, which prohibits the long-range order associated with a spontaneous breaking of a continuous symmetry, is evaded~\cite{Bruno2001,Sbierski2024}.   

Let us explain more details of the terms $\hat{H}^{(0)}$, $\hat{H}^{(1)}$, $\hat{H}^{(2)}$, and $\hat{H}^{(3)}$, respectively.
The zeroth-order term $\hat{H}^{(0)}$ is equal to the ground state energy with no fluctuation,
\begin{align}
  \hat{H}^{(0)} = M E^{\mathrm{MF}}_{\rm gs}, \label{eq:H0}
\end{align}
where $E^{\mathrm{MF}}_{\rm gs}$ is the mean-field energy per site of the $XY$-ferromagnetic ground state (See Sec.~\ref{subsec:s2a}).
The linear term $\hat{H}^{(1)}$ is given by
\begin{align}
  \hat{H}^{(1)}= A_1 (\hat{b}^{\dagger}_{\mathrm{A}, \mathbf{0}} + \hat{b}_{\mathrm{A}, \mathbf{0}}), \label{eq:H1}
\end{align}
where we have introduced the Fourier transformation of $\hat{b}^{\dagger}_{\mathrm{A}, j}$ and $\hat{b}^{\dagger}_{\Phi, j}$,
\begin{align}
  \hat{b}^{\dagger}_{m, j} = \frac{1}{\sqrt{M}} \sum_{\mathbf{k} \in \Lambda_{0}} \hat{b}^{\dagger}_{m, \mathbf{k}} e^{- i \mathbf{k} \cdot \mathbf{r}_{j}}, ~~~ m \in \left\{\mathrm{A}, \Phi \right\}. \label{FT}
\end{align}
The notation $\sum_{\mathbf{k} \in \Lambda_{0}}$ means that the momentum $\mathbf{k}$ runs over the first Brillouin zone $\Lambda_{0} \equiv [-\pi, \pi]^{d}$.
For the mean-field ground state, we can easily verify that $\hat{H}^{(1)} = 0$.
The quadratic term $\hat{H}^{(2)}$ can be written as
\begin{align}
  \hat{H}^{(2)}&= \sum_{\mathbf{k}} B_{1} \hat{b}^{\dagger}_{\mathrm{A}, \mathbf{k}} \hat{b}_{\mathrm{A}, \mathbf{k}} + \sum_{\mathbf{k}} B_{2} (\hat{b}^{\dagger}_{\mathrm{A}, \mathbf{k}} \hat{b}^{\dagger}_{\mathrm{A}, -\mathbf{k}} \nonumber \\
  &+ \hat{b}_{\mathrm{A}, \mathbf{k}} \hat{b}^{\dagger}_{\mathrm{A}, \mathbf{k}} + \hat{b}^{\dagger}_{\mathrm{A}, \mathbf{k}} \hat{b}_{\mathrm{A}, \mathbf{k}} + \hat{b}_{\mathrm{A}, \mathbf{k}} \hat{b}_{\mathrm{A}, -\mathbf{k}}) \nonumber \\
  &+ \sum_{\mathbf{k}} C_{1} \hat{b}^{\dagger}_{\Phi, \mathbf{k}} \hat{b}_{\Phi, \mathbf{k}} + \sum_{\mathbf{k}} C_{2} (\hat{b}^{\dagger}_{\Phi, \mathbf{k}} \hat{b}^{\dagger}_{\Phi, -\mathbf{k}} \nonumber \\
  &+ \hat{b}_{\Phi, \mathbf{k}} \hat{b}^{\dagger}_{\Phi, \mathbf{k}} + \hat{b}^{\dagger}_{\Phi, \mathbf{k}} \hat{b}_{\Phi, \mathbf{k}} + \hat{b}_{\Phi, \mathbf{k}} \hat{b}_{\Phi, -\mathbf{k}}). \label{eq:H2}
\end{align}
The cubic term $\hat{H}^{(3)}$, which describes the interaction between the Higgs and NG modes, can be expressed as
\begin{align}
  \hat{H}^{(3)} &= \sum_{\mathbf{k}_1, \mathbf{k}_2, \mathbf{k}_3} D_{1} (\hat{b}^{\dagger}_{\Phi, \mathbf{k}_1} \hat{b}_{\Phi, \mathbf{k}_2} \hat{b}_{\mathrm{A}, \mathbf{k}_3} + \mathrm{H.c.}) \nonumber \\
  &+ \sum_{\mathbf{k}_1, \mathbf{k}_2, \mathbf{k}_3} D_{2} (\hat{b}^{\dagger}_{\mathrm{A}, \mathbf{k}_1} \hat{b}_{\mathrm{A}, \mathbf{k}_2} \hat{b}_{\mathrm{A}, \mathbf{k}_3} + \mathrm{H.c.}) \nonumber \\
  &+ \sum_{\mathbf{k}_1, \mathbf{k}_2, \mathbf{k}_3} D_{3} (\hat{b}^{\dagger}_{\mathrm{A}, \mathbf{k}_1} \hat{b}_{\Phi, \mathbf{k}_2} \hat{b}_{\Phi, \mathbf{k}_3} \nonumber \\ 
  &- \hat{b}^{\dagger}_{\Phi, \mathbf{k}_1} \hat{b}_{\mathrm{A}, \mathbf{k}_2} \hat{b}_{\Phi, \mathbf{k}_3} + \mathrm{H.c.}). \label{eq:H3}
\end{align}
The detailed derivation of the effective Hamiltonian from Eq.~\eqref{eq:Hamiltonian}, including the inverse canonical transformation and the Holstein--Primakoff expansion up to cubic order, is presented in Appendix~\ref{App:Derivation}.
For the explicit expressions of the coefficients $A_1$, $B_1$, $B_2$, $C_1$, $C_2$, $D_1$, $D_2$, and $D_3$, see Appendix~\ref{A2}.

\subsection{Bogoliubov transformation} \label{s2e}
In Sec.~\ref{s2d}, we have obtained the series of the Hamiltonian in terms of $\hat{b}_{\mathrm{A}, j}$ and $\hat{b}_{\Phi, j}$.
As we have seen in Sec.~\ref{s2d}, $\hat{H}^{(2)}$ has no mixing term between branches labeled by $\mathrm{A}$ and $\Phi$.
Hence, we can diagonalize $\hat{H}^{(2)}$ by performing the Bogoliubov transformation independently in each branch as
\begin{align}
  \hat{b}_{m, \mathbf{k}} = u_{m, \mathbf{k}} \hat{\eta}_{m, \mathbf{k}} + v^{\ast}_{m, -\mathbf{k}} \hat{\eta}^{\dagger}_{m, -\mathbf{k}}, \label{eq:bogot}
\end{align}
where the operator $\hat{\eta}_{m,{\bf k}}$ obeys the bosonic commutation relations $[\hat{\eta}_{m, \mathbf{k}}, \hat{\eta}^{\dagger}_{n, -\mathbf{k}'}] = \delta_{m,n} \delta_{\mathbf{k}, \mathbf{k}'}$ and $[\hat{\eta}_{m, \mathbf{k}}, \hat{\eta}_{n, -\mathbf{k}'}] = [\hat{\eta}^{\dagger}_{m, \mathbf{k}}, \hat{\eta}^{\dagger}_{n, -\mathbf{k}'}] = 0$.

Let us assume that the coefficients are real and have a symmetry under a sign change of the momentum $\mathbf{k} \to -\mathbf{k}$.
Under this assumption, one can obtain the coefficients of the transformation,
\begin{align}
  u_{{\rm A}, \mathbf{k}} &= \sqrt{\frac{2 - u^2 \gamma_{\mathbf{k}}}{4 \sqrt{1 - u^{2} \gamma_{\mathbf{k}}}} + \frac{1}{2}}, \\
  v_{{\rm A}, \mathbf{k}} &= \mathrm{sgn}(\gamma_{\mathbf{k}}) \sqrt{\frac{2 - u^2 \gamma_{\mathbf{k}}}{4 \sqrt{1 - u^{2} \gamma_{\mathbf{k}}}} - \frac{1}{2}}, \\
  u_{\Phi, \mathbf{k}} &= \sqrt{\frac{2 - \gamma_{\mathbf{k}}}{4 \sqrt{1 - \gamma_{\mathbf{k}}}} + \frac{1}{2}}, \\
  v_{\Phi, \mathbf{k}} &= - \mathrm{sgn}(\gamma_{\mathbf{k}})\sqrt{\frac{2 - \gamma_{\mathbf{k}}}{4 \sqrt{1 - \gamma_{\mathbf{k}}}} - \frac{1}{2}},
\end{align}
where 
\begin{eqnarray}
\gamma_{\mathbf{k}} = \frac{1}{z_{\mathrm{eff}}}\sum_{j\neq 0} \frac{e^{i \mathbf{k} \cdot \mathbf{r}_{j}}}{|\mathbf{r}_{0,j}|^{\alpha}}.
\end{eqnarray} 
Notice that $-Jz_{\rm eff}\gamma_{\bf k}$ forms the band structure of a single particle in the hypercubic lattice.
After this transformation, the Hamiltonian of Eq.~\eqref{HPHam} reads 
\begin{align}
  \hat{H} \simeq \sum_{m \in \{\mathrm{A}, \Phi\}} \sum_{\mathbf{k}} \mathcal{E}_{m, \mathbf{k}} \hat{\eta}^{\dagger}_{m, \mathbf{k}} \hat{\eta}_{m, \mathbf{k}} + \hat{H}^{(3)}(\hat{\eta}^{\dagger}_{m, \mathbf{k}}, \hat{\eta}_{m, \mathbf{k}}). \label{eq:diagHam}
\end{align}
The dispersions of the Higgs and NG modes are given by
\begin{align}
\mathcal{E}_{\mathrm{A}, \mathbf{k}} &= 4 J z_{\mathrm{eff}} \sqrt{1 - u^{2} \gamma_{\mathbf{k}}}, 
\label{eq:higgsE} \\
\mathcal{E}_{\Phi, \mathbf{k}} &= 2J z_{\mathrm{eff}} (1 + u) \sqrt{1 - \gamma_{\mathbf{k}}}.
\label{eq:ngE}
\end{align}
The former dispersion, corresponding to the Higgs mode, has an energy gap 
\begin{eqnarray}
\Delta_{\mathrm{h}} = 4 J z_{\mathrm{eff}} \sqrt{1 - u^2}
\label{eq:HiggsGap}
\end{eqnarray} 
at $\mathbf{k} = \mathbf{0}$, whereas the latter dispersion, corresponding to the NG mode, is gapless.
The energy gap $\Delta_{\mathrm{h}}$ closes at the critical point $u = u_{\rm c} = 1$.

\subsection{Finite-temperature Green's functions} \label{s2f}
To evaluate the Beliaev damping rate of the Higgs mode, we employ the finite-temperature Green’s-function formalism \cite{Abrikosov1975, Lifshitz1980, Altland2010}.
The single-particle Green’s function of the Higgs mode is defined as
\begin{align}
  \mathcal{G}_{\mathrm{A}, \mathbf{k}}(i \omega_n) \equiv \frac{\int \mathcal{D}(\eta, \bar{\eta}) \eta_{\mathrm{A}, \mathbf{k}}(i \omega_n) \bar{\eta}_{\mathrm{A}, \mathbf{k}}(i \omega_n) \exp (- \mathcal{S})}{\int \mathcal{D}(\eta, \bar{\eta}) \exp (- \mathcal{S})}, \label{Greensfunc}
\end{align}
where $\omega_n = \frac{2 \pi n}{ \beta} \quad (n \in \mathbb{N})$ is the Matsubara frequency \cite{Abrikosov1975,Lifshitz1980,Altland2010}, $\beta = T^{-1}$ is the inverse temperature, $\eta_{\mathrm{A}, \mathbf{k}}(i\omega_n)$ and its conjugate $\bar{\eta}_{\mathrm{A}, \mathbf{k}}(i\omega_n) = [\eta_{\mathrm{A}, \mathbf{k}}(i\omega_n)]^{\ast}$ are complex-valued field variables at $(\omega_n, \mathbf{k})$, and $\mathcal{D}(\eta, \bar{\eta})$ is a measure of the integrations.
The action 
\begin{eqnarray}
\mathcal{S} = \mathcal{S}^{(2)} + \mathcal{S}^{(3)}_{{\rm damping}} + \mathcal{S}^{(3)}_{{\rm other}}
\end{eqnarray}
 is derived from Eq.~\eqref{eq:diagHam}. The quadratic action $\mathcal{S}^{(2)}$ is given by
\begin{align}
  \mathcal{S}^{(2)} \!=\! \sum_{m \in \{\mathrm{A}, \Phi\}}\!\sum_{\mathbf{k}}\sum_{n}  (- i \omega_n + \mathcal{E}_{m, \mathbf{k}}) \bar{\eta}_{m, \mathbf{k}}(i \omega_n) \eta_{m, \mathbf{k}}(i \omega_n).
\end{align}
$\mathcal{S}^{(3)}_{{\rm damping}}$ stems from a part of the cubic terms $\hat{H}^{(3)}$, which includes only the $\hat{b}^{\dagger}_{\mathrm{A}, \mathbf{k}_1} \hat{b}_{\Phi, \mathbf{k}_2} \hat{b}_{\Phi, \mathbf{k}_3}$ term and its Hermitian conjugate and is given by
\begin{align}
  \mathcal{S}^{(3)}_{{\rm damping}} = &\frac{1}{2 \sqrt{\beta}} \sum_{n_1,n_2,n_3} \sum_{\mathbf{k}_1, \mathbf{k}_2, \mathbf{k}_3}\delta_{n_1,n_2+n_3}\mathcal{M}_{\mathbf{k}_1, \mathbf{k}_2, \mathbf{k}_3} \nonumber \\
  & \times \left[ \eta_{\mathrm{A}, \mathbf{k}_1}(i \omega_{n_{1}}) \bar{\eta}_{\Phi, \mathbf{k}_2}(i \omega_{n_{2}}) \bar{\eta}_{\Phi, \mathbf{k}_3}(i \omega_{n_{3}}) + \mathrm{c.c.} \right].
\end{align}
For the explicit form of $\mathcal{M}_{\mathbf{k}_1, \mathbf{k}_2, \mathbf{k}_3}$, see Appendix~\ref{A2}. 
$\mathcal{S}^{(3)}_{{\rm other}}$ contains the other part of the cubic terms. We separately describe $\mathcal{S}^{(3)}_{{\rm damping}}$ and $\mathcal{S}^{(3)}_{{\rm other}}$ because, owing to the energy and momentum conservation laws, the former contributes to the damping of the Higgs mode with ${\bf k}={\bf 0}$ but the latter does not. 

The single-particle Green's function $\mathcal{G}_{\mathrm{A}, \mathbf{k}}(i \omega_n)$ fulfills the Dyson's equations \cite{Lifshitz1980,Altland2010}:
\begin{align}
  \mathcal{G}_{\mathrm{A}, \mathbf{k}}(i \omega_n) = \mathcal{G}^{(0)}_{\mathrm{A}, \mathbf{k}}(i \omega_n) + \mathcal{G}^{(0)}_{\mathrm{A}, \mathbf{k}}(i \omega_n) \tilde{\Sigma}_{\mathrm{A},{\bf k}}(i \omega_n) \mathcal{G}_{\mathrm{A}, \mathbf{k}}(i \omega_n), 
\end{align}
where $\tilde{\Sigma}_{\mathrm{A},{\bf k}}(i \omega_n)$ is the self-energy function of the Higgs mode and $\mathcal{G}^{(0)}_{\mathrm{A}, \mathbf{k}}(i \omega_n) = 1 / (i \omega_n - \mathcal{E}_{\mathrm{A}, \mathbf{k}})$ is the free propagator of the Higgs mode.

We evaluate the lowest-order (one-loop) self-energy diagram using the perturbative Green’s-function formalism introduced in Sec.~\ref{s2f}.
Hereafter, we focus on the Higgs mode with zero momentum $\mathbf{k} = {\bf 0}$ because such an excitation can be created relatively easily in experiments of Rydberg-atom arrays by a global control of the parameter $\Delta$. 

At second order, the self-energy of the Higgs mode with $\mathbf{k} = {\bf 0}$ is given by
\begin{align}
  \tilde{\Sigma}_{\mathrm{A},{\bf k}={\bf 0}}(i\omega_n) = \frac{1}{2}  \sum_{\mathbf{k} \in \Lambda_{0}}|\mathcal{M}_{\mathbf{0}, \mathbf{k}, -\mathbf{k}}|^2 \frac{1 + 2 n_{\mathrm{B}}\left(\mathcal{E}_{\Phi, \mathbf{k}}\right)}{i \omega_n - 2 \mathcal{E}_{\Phi, \mathbf{k}}},
  \label{SelfEnergy}
\end{align}
where $n_{\mathrm{B}}(x) = (e^{\beta x} - 1)^{-1}$ is the Bose distribution function.

In obtain the damping rate of the Higgs mode $\Gamma_{\mathrm{A}, \mathbf{k}=\mathbf{0}}$, we make an analytic continuation, $\omega_n\rightarrow\omega+i\epsilon$, that converts the self-energy for the Matsubara Green's function to the one for the retarded Green's function and take the imaginary part of the latter 
as 
\begin{align}
  \Gamma_{\mathrm{A}, \mathbf{k}=\mathbf{0}} &=\mathrm{Im}\tilde{\Sigma}_{\mathrm{A},{\bf k}={\bf 0}}(i \omega_n)|_{i \omega_n \to \omega + i \epsilon}
  \nonumber\\
&=   \frac{\pi}{2} \sum_{\mathbf{k} \in \Lambda_{0}}|\mathcal{M}_{\mathbf{0}, \mathbf{k}, -\mathbf{k}}|^2 (1 + 2 n_{\mathrm{B}}[\mathcal{E}_{\Phi, \mathbf{k}}]) \delta(\mathcal{E}_{\mathrm{A}, \mathbf{0}} - 2\mathcal{E}_{\Phi, \mathbf{k}}). \label{BeliaevDampingRate}
\end{align}
Equation~\eqref{BeliaevDampingRate} describes a Beliaev damping process, where one Higgs mode with zero momentum collapses into two NG modes with opposite momenta $\mathbf{k}$ and $-\mathbf{k}$ with satisfying the on-shell energy-momentum conservation of $\mathcal{E}_{\mathrm{A}, \mathbf{0}} = 2\mathcal{E}_{\Phi, \mathbf{k}}$.

\section{Results} \label{sec:Results}
In this section, we discuss properties of the NG and Higgs modes with the focus on their low-energy behavior.

\subsection{Excitation spectra}
When $d < \alpha < d + 2$, $\gamma_{\mathbf{k}}$ at small $|{\bf k}|$ can be expanded with respect to $|{\bf k}|$ as
\begin{eqnarray}
\gamma_{\mathbf{k}} \simeq 1 - A |\mathbf{k}|^{\alpha - d},
\label{eq:gamsmallk}
\end{eqnarray}
where $A$ is a coefficient that depends on $\alpha$ and $d$ \cite{Diessel2023,Song2023,Adelhardt2025,Yabuuchi2025}.
For the specific case of $\alpha = 3$ and $d = 2$, $A$ can be readily evaluated using Ewald summation \cite{Peter2012} as
$A=\frac{2 \pi}{z_{\mathrm{eff}}}$.
Substituting Eq.~(\ref{eq:gamsmallk}) into Eqs.~(\ref{eq:higgsE}) and (\ref{eq:ngE}), the dispersion relations of the Higgs and NG modes at small $|{\bf k}|$ are approximated as
\begin{align}
\mathcal{E}_{\mathrm{A}, \mathbf{k}} &\simeq \sqrt{\Delta_{\mathrm{h}}^{2} + g_{\mathrm{h}}^{2} |\mathbf{k}|^{\alpha-d}}
\simeq \Delta_{\rm h} + \frac{g_{\rm h}^2}{2\Delta_{\rm h}}|\mathbf{k}|^{\alpha-d},
\label{eq:Higgs_dispersion}
\\
\mathcal{E}_{\Phi, \mathbf{k}} &\simeq g_{\mathrm{ng}} |\mathbf{k}|^{\frac{\alpha-d}{2}}, \label{eq:NG_dispersion}
\end{align}
where 
\begin{eqnarray}
g_{\mathrm{h}}&=&4Jz_{\rm eff}u\sqrt{A},
\\
g_{\mathrm{ng}}&=&2Jz_{\rm eff}(1+u)\sqrt{A}.
\end{eqnarray}
Since in the case of the nearest-neighbor interaction $(\mathcal{E}_{\mathrm{A}, \mathbf{k}}-\Delta_{\rm h})\propto |{\bf k}|^2$ and $\mathcal{E}_{\Phi, \mathbf{k}}\propto |{\bf k}|$~\cite{Altman2002,Nagao2016,Nagao2016b}, Eqs.~(\ref{eq:Higgs_dispersion}) and (\ref{eq:NG_dispersion}) mean that the long-range nature of the spin-exchange interactions significantly modifies the dispersion relations. In the limit of $\alpha\rightarrow d+2$, the qualitative behavior of the dispersion relations at small $|{\bf k}|$ agrees with that for the nearest-neighbor interaction. 
\begin{figure}[t]
\centering
\includegraphics[width=\columnwidth]{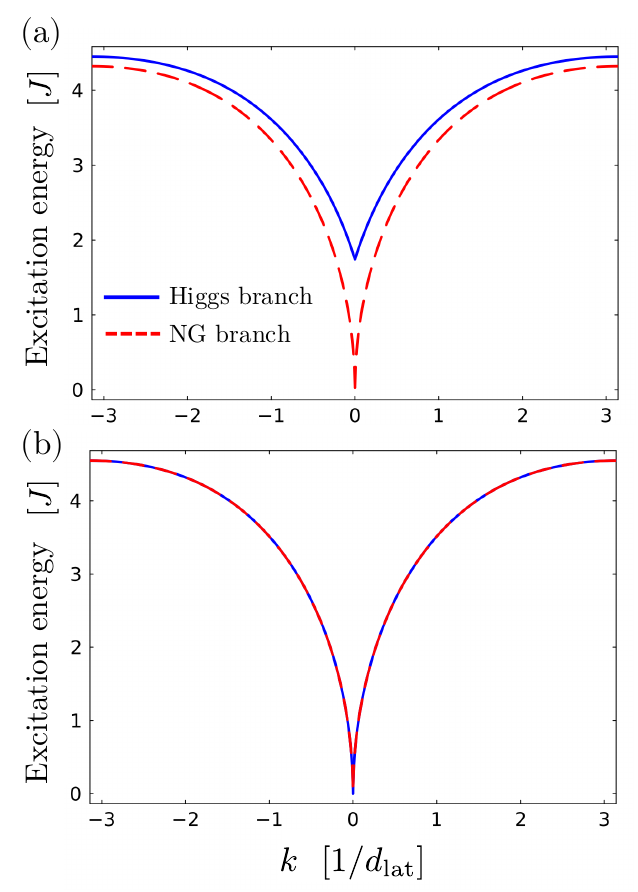}
\caption{Dispersion relations of the Higgs (blue solid line) and NG (red dashed line) modes, which are expressed in Eqs.~(\ref{eq:higgsE}) and (\ref{eq:ngE}), for $d=2$, $\alpha=3$, and ${\bf k}=(k,0)$, where (a) $u =0.9(< u_{\rm c})$ and (b) $u = u_{\rm c}$. }
\label{fig:dispersion}
\end{figure}

\begin{figure}[t]
\centering
\includegraphics[width=\columnwidth]{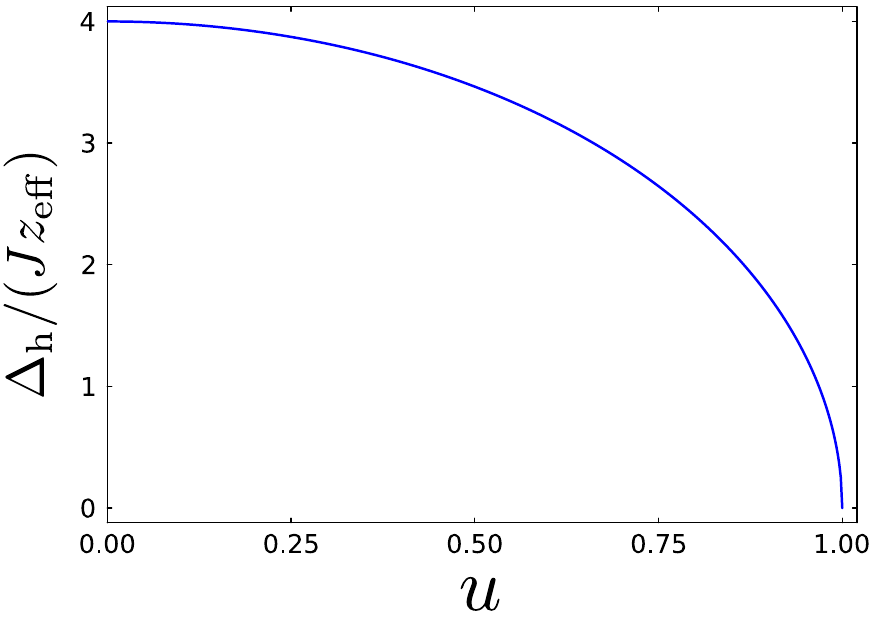}
\caption{Higgs gap $\Delta_{\rm h}$ as a function of $u$, whose analytical expression is given in Eq.~(\ref{eq:HiggsGap}).}
\label{fig:HiggsGap}
\end{figure}

In Fig.~\ref{fig:dispersion}, we show the dispersion relations $\mathcal{E}_{\mathrm{A}, \mathbf{k}}$ and $\mathcal{E}_{\Phi, \mathbf{k}}$ of Eqs.~(\ref{eq:higgsE}) and (\ref{eq:ngE}) for $\alpha=3$ and $d=2$ by setting ${\bf k}=(k,0)$. When $u=0.9$ [Fig.~\ref{fig:dispersion}(a)], the Higgs mode exhibits a gapped and linear dispersion, while the NG mode exhibits a gapless square root dispersion at small $|{\bf k}|$.
As shown in Fig.~\ref{fig:dispersion}(b), at the critical point $u=1$, the gap of the Higgs mode vanishes such that the dispersion of the Higgs mode coincides with that of the NG mode. In Fig.~\ref{fig:HiggsGap}, we show the gap of the Higgs mode $\Delta_{\rm h}$ as a function of $u$. There we see that the Higgs gap obeys the critical scaling behavior $\Delta_{\mathrm{h}} \sim |u - u_\mathrm{c}|^{\frac{1}{2}}$. While this gap closing also occurs in the case of the nearest-neighbor interactions, the critical exponent deviates from the mean-field value $\frac{1}{2}$~\cite{Pollet2012}.

\subsection{Beliaev damping rate}
We discuss the Beliaev damping of the Higgs mode at finite temperatures by obtaining an approximate analytical expression of the damping rate $\Gamma_{{\rm A},{\bf k}={\bf 0}}$.
For this purpose, let us evaluate the integrals of Eq.~\eqref{BeliaevDampingRate} within the long-wavelength approximation, where $u_{m, \mathbf{k}}$ and $v_{m, \mathbf{k}}$ are approximated as
\begin{align*}
  u_{\mathrm{A}, \mathbf{k}} \simeq \frac{1}{\sqrt{2 \bar{\Delta}_{\mathrm{h}}}}, &\quad v_{\mathrm{A}, \mathbf{k}} \simeq \frac{1}{\sqrt{2 \bar{\Delta}_{\mathrm{h}}}}, \\
  u_{\Phi, \mathbf{k}} \simeq \sqrt{\frac{u + 1}{4 \bar{g}_{\mathrm{ng}} |\mathbf{k}|^{\frac{\alpha - d}{2}}}}, &\quad v_{\Phi, \mathbf{k}} \simeq -\sqrt{\frac{u + 1}{4 \bar{g}_{\mathrm{ng}} |\mathbf{k}|^{\frac{\alpha - d}{2}}}},
\end{align*}
where $\bar{\Delta}_{\mathrm{h}} = \frac{\Delta_{\mathrm{h}}}{J z_{\mathrm{eff}}}$ and $\bar{g}_{\mathrm{ng}} = \frac{g_{\mathrm{ng}}}{J z_{\mathrm{eff}}}$.
This approximation is better justified in a closer vicinity of the critical point $u = u_{\rm c}$, where the momenta of the NG modes with energy $\frac{\Delta_{\mathrm{h}}}{2}$ contributed dominantly to the damping of the Higgs mode are smaller.
Performing the integration with respect to $\mathbf{k}$ on the right hand side of Eq.~\eqref{BeliaevDampingRate}, we obtain the simple formula,
\begin{align}
  \Gamma_{\mathrm{A}, \mathbf{k}=\mathbf{0}} \simeq &\frac{\pi J z_{\mathrm{eff}} }{8 (\alpha - d)} \frac{\Omega_d \sqrt{1 - u^2} (1 + u)}{(2\pi)^d A^{3/2}} \nonumber \\
  &\times \left(\frac{\Delta_{\mathrm{h}}}{2 g_{\mathrm{ng}}}\right) ^{\frac{5d - 3 \alpha}{\alpha - d}} \coth \left( \frac{\beta \Delta_{\mathrm{h}}}{4} \right), \label{IntegratingDampingRate}
\end{align}
where 
\begin{eqnarray}
\Omega_d = \frac{2 \pi^{\frac{d}{2}}}{\Gamma(\frac{d}{2})}
\end{eqnarray}
denotes the surface area of a unit sphere in $d$ dimensions.
The intermediate steps of the above integration are presented in Appendix~\ref{App:Damping}.
It is obvious from Eq.~\eqref{IntegratingDampingRate} that the dependence on temperature $T$ is determined by the factor $\coth\left(\frac{\beta \Delta_{\mathrm{h}}}{4}\right)$.

For the specific case of $\alpha = 3$ and $d = 2$, the damping rate becomes
\begin{align}
  \Gamma_{\mathrm{A}, \mathbf{k}={\bf 0}} = \frac{J z_{\mathrm{eff}}^3 (1 - u^2)}{16 \pi^2}  \coth\left( \frac{\beta \Delta_{\mathrm{h}}}{4} \right). 
\label{eq:DipolarDampingRate}
\end{align}
For another specific case of $\alpha \to 5$ and $d = 3$, which corresponds to previous works \cite{Altman2002, Nagao2016},  it is
\begin{align}
  \Gamma_{\mathrm{A}, \mathbf{k}={\bf 0}} = \frac{3^{\frac{3}{2}}J z_{\mathrm{eff}} (1+u)\sqrt{1 - u^2}}{2\sqrt{2} \pi}  \coth\left( \frac{\beta \Delta_{\mathrm{h}}}{4} \right). 
\label{eq:DampingRateNN}
\end{align}
The expression of Eq.~(\ref{eq:DampingRateNN}) with the replacement of $z_{\rm eff}$ by the coordination number $z$ agrees with those in the previous works.

\begin{figure}[t]
  \centering
  \includegraphics[width=\columnwidth]{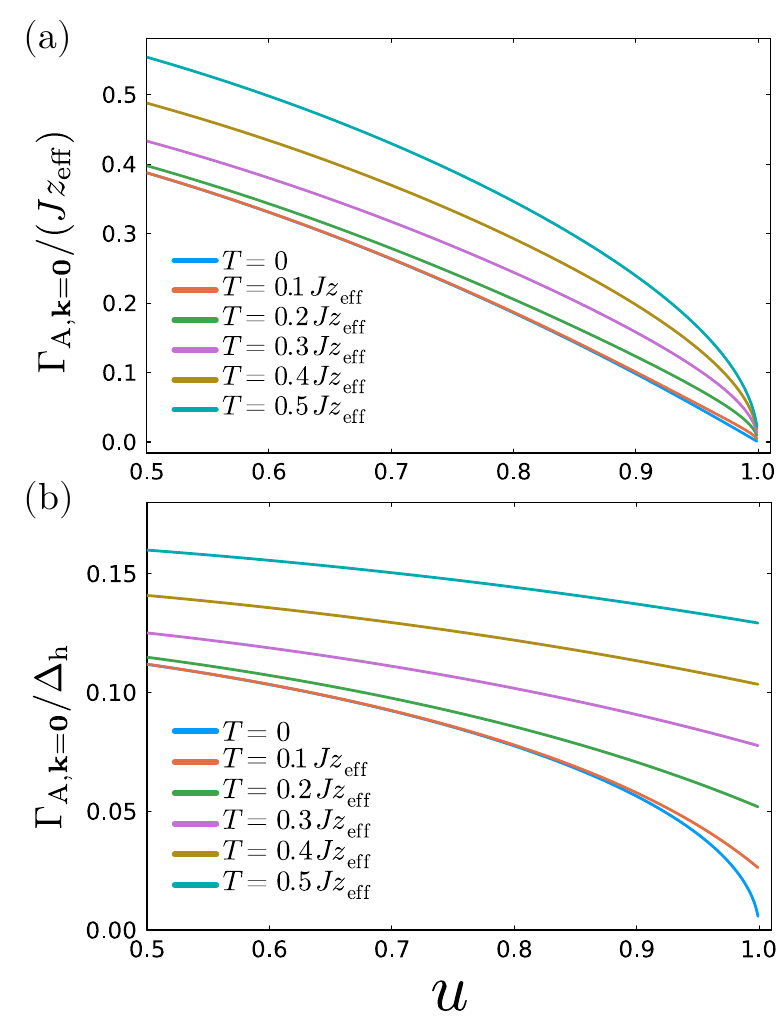}
  \caption{Beliaev damping rate of the Higgs mode as a function of $u$ for $d = 2$ and $\alpha = 3$, which is expressed in Eq.~(\ref{eq:DipolarDampingRate}). (a) $\Gamma_{\mathrm{A}, \mathbf{k} = \mathbf{0}} / (J z_{\mathrm{eff}})$ and (b) $\Gamma_{\mathrm{A}, \mathbf{k} = \mathbf{0}} / \Delta_{\mathrm{h}}$.}
  \label{fig:damping}
\end{figure}

In Fig.~\ref{fig:damping}(a), we show the damping rate $\frac{\Gamma_{{\rm A},{\bf k}={\bf 0}}}{Jz_{\rm eff}}$ as a function of $u$ for $\alpha=3$, $d=2$, and several values of the temperature.
We find that the damping rate increases monotonically with the temperature for a given $u$, but it vanishes at the critical point regardless of the value of the temperature.
In Fig.~\ref{fig:damping}(b), we show the damping rate measured by the Higgs gap $\frac{\Gamma_{{\rm A},{\bf k}={\bf 0}}}{\Delta_{\rm h}}$ for $\alpha=3$, $d=2$, and several values of the temperature. At a finite temperature in the region of $0<T\leq 0.5Jz_{\rm eff}$, $\frac{\Gamma_{{\rm A},{\bf k}={\bf 0}}}{\Delta_{\rm h}}$ converges to a finite value, which is sufficiently smaller than unity, in the limit of $u\rightarrow 1$. This fact indicates that the long-range interaction significantly suppresses the damping of the Higgs mode near the critical point so that the collective oscillation of the Higgs mode can be long-lived at sufficiently low temperatures, i.e., not over-damped at least. Notice that a similar suppression of the damping of the Higgs mode due to a kind of long-range interaction has been pointed out for a two-dimensional superconductor~\cite{Gao2021}. Nevertheless, the mechanism of the suppression is totally different from our case in the sense that there the type of long-range interaction is the density-density interaction and the ordered phase is a fermionic superconductor with Cooper pairs.

It is worth emphasizing that in the case of the nearest-neighbor interaction, the collective oscillation of the Higgs mode at ${\bf k}={\bf 0}$ is over-damped at finite temperatures even for $d=3$~\cite{Nagao2016,Nagao2016b}. We can understand this from Eq. (\ref{eq:DampingRateNN}). Specifically, $\frac{\Gamma_{{\rm A},{\bf k}={\bf 0}}}{\Delta_{\rm h}}$ for $\alpha \to 5$ and $d = 3$ is proportional to $\frac{T}{\Delta_{\rm h}}$ near the critical point so that it diverges in the limit of $u\rightarrow 1$.
By contrast, Eq.~(\ref{eq:DipolarDampingRate}) implies that $\frac{\Gamma_{{\rm A},{\bf k}={\bf 0}}}{\Delta_{\rm h}}$ for $\alpha = 3$ and $d = 2$ is proportional to $T$ so that it converges in the limit of $u\rightarrow 1$.  

We finally discuss how to create and observe the Higgs mode at ${\bf k}={\bf 0}$ in experiments of Rydberg-atom arrays described by Eq.~(\ref{eq:Hamiltonian}). To be specific, we consider the case in which the local spin-1 states are formed by the $|41F\rangle$, $|43D\rangle$, and $|45P\rangle$ Rydberg states as in Ref.~\cite{Chew2022}. Recall that $\Delta$ can be dynamically controlled in experiments by changing the light shift of one of the three states~\cite{Ravets2014}. One needs to prepare the $XY$-ferromagnetic ordered state with $\bar{S}^z=0$ that possesses the Higgs mode as an elementary excitation. This can be achieved as follows: (i) set $\Delta$ to be much larger than $Jz_{\rm eff}$ and excite all the atoms from the ground state to $|43D\rangle$($=|S^z=0\rangle$) such that the system is prepared to be in the disordered state with $\bar{S}^z=0$ the ground state for the given $\frac{\Delta}{Jz_{\rm eff}}$. (ii) Decrease $\Delta$ slowly down across the critical point $u=1$ and stop the decrease around $u\simeq 1$, e.g., $u=0.9$ in order to prepare the $XY$-ferromagnetic ordered state. From this state, (iii) induce the oscillation of the order-parameter amplitude, i.e., the Higgs mode by suddenly quenching $\Delta$, e.g., from $0.9$ to $0.95$. For the detection of the Higgs mode, (iv) measure the time evolution of the occupation probability of each local state as done in Refs.~\cite{Chew2022}. The occupation probability should oscillate with the same frequency and damping rate as those of the Higgs mode. Thus, we suggest that the Higgs mode can be created and detected through procedure (i)--(iv). The temporal oscillations of the order parameter have been observed for Rydberg atom arrays described by the spin-1/2 antiferromagnetic Ising model with a mixed field, whose ground-state can be the antiferromagnetic ordered phase breaking spontaneously a discrete symmetry, in a similar setting that utilizes the combination of an adiabatic preparation of the initial state and a sudden quench of a Hamiltonian parameter~\cite{Manovitz2025}. Moreover, a similar quench setting for probing the Higgs mode has been proposed in the context of fermionic superfluids~\cite{Barankov2004,Yuzbashyan2006,Tokimoto2019,Collado2023}.

\section{Conclusion} \label{sec:Conclusion}
By means of the mean-field and quantum-field-theoretical approaches, we have analytically investigated the collective excitations in the $XY$-ferromagnetic ordered phase of the spin-1 $XY$ model with a quadratic Zeeman term and the long-range interaction that decays with distance $r$ as $\propto r^{-\alpha}$.
For the specific case of $\alpha=3$ and two dimensions, which corresponds to Rydberg-atom experiments, we found that the excitation energy of the Higgs mode exhibits a linear dispersion with a finite-energy gap, whereas the dispersion of the NG mode becomes proportional to the square root of the momentum.
Furthermore, our analysis revealed that long-range  interaction significantly suppresses the Beliaev damping of the Higgs mode near the critical point to the disordered phase.
These results suggest that the Higgs mode can be detected as a long-lived oscillation of the order parameter. We proposed a specific protocol to experimentally create and probe the Higgs mode as such an oscillation. 

In the present paper, we focused on the homogeneous system. However, given the high controllability of the spatial configuration of Rydberg-atom arrays, it will be interesting to explore effects of the spatial inhomogeneity on the Higgs mode. One interesting possibility is the study of the Higgs mode in the ferromagnetic phase of the spin-$1/2$ $XY$ model under a staggered magnetic field~\cite{Chen2023} near the quantum phase transition to the disordered phase. It will be also interesting to explore Higgs bound states in the presence of local distortions of the spin-exchange interactions that have been theoretically predicted for the case of the short-range interaction~\cite{Nakayama2015,Danshita2017}.
\begin{acknowledgments}
We thank Y. Kondo, K. Nagao, and T. Tomita for useful discussions.
This work was supported by JST ASPIRE [Grant No.~JPMJAP24C2], MEXT Quantum Leap Flagship Program (MEXT Q-LEAP) [Grant No.~JPMXS0118069021], JST FOREST [Grant No.~JPMJFR202T], and JSPS KAKENHI [Grant No.~JP26K00639].
\end{acknowledgments}

\appendix
\section{Supplement on the derivation of the effective Hamiltonian} \label{App:Derivation}
In this Appendix, we derive the effective Hamiltonian used in Sec.~\ref{sec:Methods} from the Hamiltonian in Eq.~\eqref{eq:Hamiltonian}. Our goal is to make explicit the intermediate steps leading to Eqs.~\eqref{eq:H0}, \eqref{eq:H1}, \eqref{eq:H2}, \eqref{eq:H3}, and \eqref{eq:diagHam}.

The derivation proceeds as follows. First, we rewrite the Schwinger bosons in terms of the rotated bosons $\hat{b}_{0,j}$, $\hat{b}_{{\rm A},j}$, and $\hat{b}_{\Phi,j}$, where $\hat{b}^{\dagger}_{0,j}$ creates the local mean-field ground states while $\hat{b}^{\dagger}_{{\rm A},j}$ and $\hat{b}^{\dagger}_{\Phi,j}$ create states corresponding to the amplitude and phase fluctuations, respectively. Next, using the local constraint, we eliminate $\hat{b}_{0,j}$ and perform the Holstein--Primakoff expansion up to third order in the fluctuation operators. This is the lowest order relevant to the Beliaev damping process of the Higgs mode. Finally, after Fourier and Bogoliubov transformations, we obtain the effective Hamiltonian that describes excitation spectra and interactions between NG and Higgs modes.

We begin by inverting the canonical transformation in Eq.~\eqref{CnT}. This allows us to express the original spin operators and the Hamiltonian directly in terms of the fluctuation operators around the mean-field ground state:
\begin{align}
  \hat{t}_{0,j} &= c_1 \hat{b}_{0,j} + s_1 \hat{b}_{{\rm A},j}, \nonumber \\
  \hat{t}_{1,j} &= \frac{1}{\sqrt{2}}\left(s_1 \hat{b}_{0,j} - c_1 \hat{b}_{{\rm A},j} + \hat{b}_{\Phi,j}\right), \label{eq:InCnT} \\
  \hat{t}_{-1,j} &= \frac{1}{\sqrt{2}}\left(s_1 \hat{b}_{0,j} - c_1 \hat{b}_{{\rm A},j} - \hat{b}_{\Phi,j}\right), \nonumber 
\end{align}
where the coefficients are $s_{1} = \sin\left(\frac{\theta_{\mathrm{gs}}}{2}\right)$ and $c_{1} = \cos\left(\frac{\theta_{\mathrm{gs}}}{2}\right)$.

Using Eq.~\eqref{eq:InCnT}, the spin operators in the Schwinger-boson representation can be rewritten in terms of the new bosons $\hat{b}_{0, j}$, $\hat{b}_{{\rm A}, j}$, and $\hat{b}_{\Phi, j}$ in the following manner:
\begin{align}
  \hat{S}^{+}_{j} &=2 s_{1}c_{1} \hat{b}^{\dagger}_{0, j} \hat{b}_{0, j} - 2 s_{1}c_{1} \hat{b}^{\dagger}_{{\rm A}, j} \hat{b}_{{\rm A}, j} \nonumber\\
  &+ (s^{2}_{1} - c^{2}_{1})(\hat{b}^{\dagger}_{{\rm A}, j} \hat{b}_{0, j} + \hat{b}^{\dagger}_{0, j} \hat{b}_{{\rm A}, j}) \nonumber\\
  &- s_{1}(\hat{b}^{\dagger}_{{\rm A}, j} \hat{b}_{\Phi, j} - \hat{b}^{\dagger}_{\Phi, j} \hat{b}_{{\rm A}, j}) - c_{1}(\hat{b}^{\dagger}_{0, j} \hat{b}_{\Phi, j} - \hat{b}^{\dagger}_{\Phi, j} \hat{b}_{0, j}), \label{eq:Sp_b}\\
  \hat{S}_{j}^{z} &= s_{1} \left( \hat{b}_{0,j}^{\dagger}\hat{b}_{\Phi,j} + \hat{b}_{\Phi,j}^{\dagger}\hat{b}_{0,j} \right) - c_{1}\left( \hat{b}_{{\rm A},j}^{\dagger}\hat{b}_{\Phi,j} + \hat{b}_{\Phi,j}^{\dagger}\hat{b}_{{\rm A},j} \right), \label{eq:Sz_b} \\
  (\hat{S}_{j}^{z})^{2} &= s_{1}^{2}\hat{b}_{0,j}^{\dagger}\hat{b}_{0,j} + c_{1}^{2}\hat{b}_{{\rm A},j}^{\dagger}\hat{b}_{{\rm A},j} + \hat{b}_{\Phi,j}^{\dagger}\hat{b}_{\Phi,j} \nonumber \\
  &- s_{1}c_{1}\left( \hat{b}_{0,j}^{\dagger}\hat{b}_{{\rm A},j} + \hat{b}_{{\rm A},j}^{\dagger}\hat{b}_{0,j} \right). \label{eq:Sz2_b}
\end{align}
Substituting these expressions into Eq.~\eqref{eq:Hamiltonian}, one can obtain the Hamiltonian in terms of the bosons $\hat{b}_{0, j}$, $\hat{b}_{{\rm A}, j}$, and $\hat{b}_{\Phi, j}$.
The Hamiltonian can be written explicitly as
\begin{widetext}
\begin{align}
  \hat{H} &= - \sum_{j<l} 8 J_{j,l} s_{1}^{2} c_{1}^{2} \hat{b}^{\dagger}_{0, j} \hat{b}_{0, j} \hat{b}^{\dagger}_{0, l} \hat{b}_{0, l} + \sum_{j<l} 16 J_{j,l} s_{1}^{2} c_{1}^{2} \hat{b}^{\dagger}_{0, j} \hat{b}_{0, j} \hat{b}^{\dagger}_{{\rm A}, l} \hat{b}_{{\rm A}, l} - \sum_{j<l} 8 J_{j,l} s_{1}^{2} c_{1}^{2} \hat{b}^{\dagger}_{{\rm A}, j} \hat{b}_{{\rm A}, j} \hat{b}^{\dagger}_{{\rm A}, l} \hat{b}_{{\rm A}, l} \nonumber \\
  &- \sum_{j<l} 2 J_{j,l} (c_{1}^{2} - s_{1}^{2})^{2} (\hat{b}^{\dagger}_{{\rm A}, j} \hat{b}_{0 ,j} \hat{b}^{\dagger}_{{\rm A}, l} \hat{b}_{0 ,l} + \hat{b}^{\dagger}_{0, j} \hat{b}_{{\rm A}, j} \hat{b}^{\dagger}_{{\rm A}, l} \hat{b}_{0, l} + \hat{b}^{\dagger}_{{\rm A}, j} \hat{b}_{0, j} \hat{b}^{\dagger}_{0, l} \hat{b}_{{\rm A}, l} + \hat{b}^{\dagger}_{0, j} \hat{b}_{{\rm A}, j} \hat{b}^{\dagger}_{0, l} \hat{b}_{{\rm A}, l}) \nonumber \\
  &+ \sum_{j<l} 2 J_{j,l} c_{1}^{2} (\hat{b}^{\dagger}_{0, j} \hat{b}_{\Phi, j} \hat{b}^{\dagger}_{0, l} \hat{b}_{\Phi, l} - \hat{b}^{\dagger}_{\Phi, j} \hat{b}_{0, j} \hat{b}^{\dagger}_{0, l} \hat{b}_{\Phi, l} + \hat{b}^{\dagger}_{\Phi, j} \hat{b}_{0, j} \hat{b}^{\dagger}_{\Phi, l} \hat{b}_{0, l} - \hat{b}^{\dagger}_{0, j} \hat{b}_{\Phi, j} \hat{b}^{\dagger}_{\Phi, l} \hat{b}_{0, l}) \nonumber \\
  &+ \sum_{j<l} 2 J_{j,l} s_{1}^{2} (\hat{b}^{\dagger}_{{\rm A}, j} \hat{b}_{\Phi, j} \hat{b}^{\dagger}_{{\rm A}, l} \hat{b}_{\Phi, l} - \hat{b}^{\dagger}_{\Phi, j} \hat{b}_{{\rm A}, j} \hat{b}^{\dagger}_{{\rm A}, l} \hat{b}_{\Phi, l} + \hat{b}^{\dagger}_{\Phi, j} \hat{b}_{{\rm A}, j} \hat{b}^{\dagger}_{\Phi, l} \hat{b}_{{\rm A}, l} - \hat{b}^{\dagger}_{{\rm A}, j} \hat{b}_{\Phi, j} \hat{b}^{\dagger}_{\Phi, l} \hat{b}_{{\rm A}, l}) \nonumber \\
  &- \sum_{j<l} 8 J_{j,l} s_{1} c_{1} (c_{1}^{2} - s_{1}^{2}) (\hat{b}^{\dagger}_{{\rm A}, j} \hat{b}_{{\rm A}, j} \hat{b}^{\dagger}_{{\rm A}, l} \hat{b}_{0, l} + \hat{b}^{\dagger}_{{\rm A}, j} \hat{b}_{{\rm A}, j} \hat{b}^{\dagger}_{0, l} \hat{b}_{{\rm A}, l} - \hat{b}^{\dagger}_{0, j} \hat{b}_{0, j} \hat{b}^{\dagger}_{{\rm A}, l} \hat{b}_{0, l} - \hat{b}^{\dagger}_{0, j} \hat{b}_{0, j} \hat{b}^{\dagger}_{0, l} \hat{b}_{{\rm A}, l}) \nonumber \\
  &+ \sum_{j<l} 4 J_{j,l} s_{1} c_{1} (\hat{b}^{\dagger}_{{\rm A}, j} \hat{b}_{\Phi, j} \hat{b}^{\dagger}_{0, l} \hat{b}_{\Phi, l} - \hat{b}^{\dagger}_{{\rm A}, j} \hat{b}_{\Phi, j} \hat{b}^{\dagger}_{\Phi, l} \hat{b}_{0, l} + \hat{b}^{\dagger}_{\Phi, j} \hat{b}_{{\rm A}, j} \hat{b}^{\dagger}_{\Phi, l} \hat{b}_{0, l} - \hat{b}^{\dagger}_{\Phi, j} \hat{b}_{{\rm A}, j} \hat{b}^{\dagger}_{0, l} \hat{b}_{\Phi, l}) \nonumber \\
  &+ \Delta \sum_{j} \left[ (c_{1}^{2} - s_{1}^{2}) \hat{b}^{\dagger}_{{\rm A}, j} \hat{b}_{{\rm A}, j} + c_{1}^{2} \hat{b}^{\dagger}_{\Phi, j} \hat{b}_{\Phi, j} - s_{1} c_{1} (\hat{b}^{\dagger}_{{\rm A}, j} \hat{b}_{0, j} + \hat{b}^{\dagger}_{0, j} \hat{b}_{{\rm A}, j}) + s_{1}^{2} \right] \nonumber \\
  &+ p \sum_{j} \left[c_{1} (\hat{b}^{\dagger}_{{\rm A}, j} \hat{b}_{\Phi, j} + \hat{b}^{\dagger}_{\Phi, j} \hat{b}_{{\rm A}, j}) - s_{1} (\hat{b}^{\dagger}_{0, j} \hat{b}_{\Phi, j} + \hat{b}^{\dagger}_{\Phi, j} \hat{b}_{0 ,j}) \right]. \label{eq:bHamiltonian}
\end{align}

In the following, we focus on the particle-hole-symmetric case and set $p=0$ as assumed in the main text. Under this condition, the phase and amplitude modes decouple already at the quadratic order, which allows us to identify the Higgs and NG branches separately.
Let us assume that the mean-field approximation is adequate for describing the $XY$-ferromagnetic ordered phase in the vicinity of the quantum phase transition to the disordered phase, so that the fluctuations described by $\hat{b}_{{\rm A},j}$ and $\hat{b}_{\Phi,j}$ remain small.
We then simplify the Hamiltonian \eqref{eq:bHamiltonian} by means of the Holstein-Primakoff expansion \eqref{HPex}.
Eliminating $\hat{b}^{\dagger}_{0, j}$ and $\hat{b}_{0, j}$ in the Hamiltonian \eqref{eq:bHamiltonian} by using the constraint \eqref{LC2}, and substituting the Holstein--Primakoff expansion \eqref{HPex} into the Hamiltonian \eqref{eq:bHamiltonian}, we can obtain the following series:
\begin{align*}
  \hat{H} = \hat{H}^{(0)} + \hat{H}^{(1)} + \hat{H}^{(2)} + \hat{H}^{(3)} + \cdots,
\end{align*}
where $\hat{H}^{(l)}$ $(l=0,1,2,3, \ldots)$ contains terms of order $l$ in the fluctuation operator $\hat{b}^{\dagger}_{{\rm A}, j}$, $\hat{b}_{{\rm A}, j}$, $\hat{b}^{\dagger}_{\Phi, j}$, and $\hat{b}_{\Phi, j}$.
Retaining terms up to cubic order is sufficient for our purpose, because the Beliaev damping of the Higgs mode is generated by the coupling between one Higgs mode and two NG modes.
The explicit forms are given by
\begin{align}
  \hat{H}^{(0)} &= \sum_{j} \left( \Delta s_{1}^{2} - 4 J z_{{\rm eff}} s_{1}^{2} c_{1}^{2} \right) = M E_{{\rm gs}}^{{\rm MF}}, \\
  \hat{H}^{(1)} &= \sum_{j} 4 J z_{{\rm eff}} s_{1} c_{1} \left( c_{1}^{2} - s_{1}^{2} - \frac{\Delta}{4 J z_{{\rm eff}}} \right) (\hat{b}^{\dagger}_{{\rm A}, j} + \hat{b}_{{\rm A}, j}), \\
  \hat{H}^{(2)} &= \sum_{j} 4 J z_{{\rm eff}} \left( 4 s_{1}^{2} c_{1}^{2} + \frac{\Delta}{4 J z_{{\rm eff}}} (c_{1}^{2} - s_{1}^{2}) \right) \hat{b}^{\dagger}_{{\rm A}, j} \hat{b}_{{\rm A}, j} + \sum_{j} 4 J z_{{\rm eff}} \left( 2 s_{1}^{2} c_{1}^{2} + \frac{\Delta}{4 J z_{{\rm eff}}} c_{1}^{2} \right) \hat{b}^{\dagger}_{\Phi, j} \hat{b}_{\Phi, j} \nonumber \\
  &- \sum_{j<l} 2 J_{j,l} (c_{1}^{2} - s_{1}^{2})^{2} (\hat{b}^{\dagger}_{{\rm A}, j} \hat{b}^{\dagger}_{{\rm A}, l} + \hat{b}_{{\rm A}, j} \hat{b}^{\dagger}_{{\rm A}, l} + \hat{b}^{\dagger}_{{\rm A}, j} \hat{b}_{{\rm A}, l} + \hat{b}_{{\rm A}, j} \hat{b}_{{\rm A}, l})\nonumber \\
  &+ \sum_{j<l} 2 J_{j,l} c_{1}^{2} (\hat{b}_{\Phi, j} \hat{b}_{\Phi, l} - \hat{b}_{\Phi, j} \hat{b}^{\dagger}_{\Phi, l} + \hat{b}^{\dagger}_{\Phi, j} \hat{b}^{\dagger}_{\Phi, l} - \hat{b}^{\dagger}_{\Phi, j} \hat{b}_{\Phi, l}), \\
  \hat{H}^{(3)} &= \sum_{j} 2 J z_{{\rm eff}} s_{1} c_{1} \left( c_{1}^{2} - s_{1}^{2} - \frac{\Delta}{4 J z_{{\rm eff}}} \right) (\hat{b}^{\dagger}_{{\rm A}, j} \hat{b}^{\dagger}_{{\rm A}, j} \hat{b}_{{\rm A}, j} + \hat{b}^{\dagger}_{{\rm A}, j} \hat{b}_{{\rm A}, j} \hat{b}_{{\rm A}, j} + \hat{b}^{\dagger}_{{\rm A}, j} \hat{b}^{\dagger}_{\Phi, j} \hat{b}_{\Phi, j} + \hat{b}_{{\rm A}, j} \hat{b}^{\dagger}_{\Phi, j} \hat{b}_{\Phi, j}) \nonumber \\
  &- \sum_{j<l} 16 J_{j,l} s_{1} c_{1} (c_{1}^{2} - s_{1}^{2}) (\hat{b}^{\dagger}_{{\rm A}, j} \hat{b}_{{\rm A}, j} \hat{b}^{\dagger}_{{\rm A}, l} + \hat{b}^{\dagger}_{{\rm A}, j} \hat{b}_{{\rm A}, j} \hat{b}_{{\rm A}, l}) - \sum_{j<l} 8 J_{j,l} s_{1} c_{1} (c_{1}^{2} - s_{1}^{2}) (\hat{b}^{\dagger}_{\Phi, j} \hat{b}_{\Phi, j} \hat{b}^{\dagger}_{{\rm A}, l} + \hat{b}^{\dagger}_{\Phi, j} \hat{b}_{\Phi, j} \hat{b}_{{\rm A}, l}) \nonumber \\
  &+ \sum_{j<l} 4 J_{j,l} s_{1} c_{1} (\hat{b}^{\dagger}_{{\rm A}, j} \hat{b}_{\Phi, j} \hat{b}_{\Phi, l} - \hat{b}^{\dagger}_{{\rm A}, j} \hat{b}_{\Phi, j} \hat{b}^{\dagger}_{\Phi, l} - \hat{b}^{\dagger}_{\Phi, j} \hat{b}_{{\rm A}, j} \hat{b}_{\Phi, l} + \hat{b}^{\dagger}_{\Phi, j} \hat{b}_{{\rm A}, j} \hat{b}^{\dagger}_{\Phi, l}).
\end{align}
The zeroth-order term reproduces the mean-field ground-state energy. The linear term vanishes when the saddle-point condition is satisfied, as expected for the expansion around the mean-field ground state. The quadratic term determines the excitations, namely the Higgs and NG modes, whereas the cubic term describes their leading interaction processes. In particular, it includes the coupling of one Higgs mode to two NG modes, which gives rise to the Beliaev damping discussed in Sec.~\ref{s2f}.

Applying the Fourier transformation \eqref{FT}, the translational invariance of the homogeneous system allows us to express the momentum dependence of the quadratic and cubic terms through $\gamma_{\mathbf{k}}$ and momentum-conservation Kronecker deltas. Collecting the resulting terms, we obtain Eqs.~\eqref{eq:H1}, \eqref{eq:H2}, and \eqref{eq:H3}.

Next, we diagonalize the quadratic Hamiltonian by means of the Bogoliubov transformation \eqref{eq:bogot}. The Bogoliubov coefficients are chosen so that the bosonic commutation relations are preserved, namely $|u_{m, \mathbf{k}}|^{2} - |v_{m, -\mathbf{k}}|^{2} = 1$. In this way, the quadratic Hamiltonian reads
\begin{align*}
  \hat{H}^{(2)} = \sum_{m \in \{{\rm A},\Phi\}} \sum_{{\bf k}} \mathcal{E}_{m, \mathbf{k}} \hat{\eta}^{\dagger}_{m, \mathbf{k}} \hat{\eta}_{m, \mathbf{k}},
\end{align*}
where $\mathcal{E}_{m, \mathbf{k}}$ is the dispersion of each branch.
The cubic Hamiltonian can also be rewritten in terms of the Bogoliubov quasiparticles $\hat{\eta}_{m,\mathbf{k}}$ as a sum of interaction terms among the Higgs and NG modes.
The explicit form is given by
\begin{align}
  \hat{H}^{(3)}(\hat{\eta}^{\dagger}_{m, \mathbf{k}}, \hat{\eta}_{m, \mathbf{k}}) = \sum_{\mathbf{k}_{1}, \mathbf{k}_{2}, \mathbf{k}_{3}} \mathcal{M}_{\mathbf{k}_{1}, \mathbf{k}_{2}, \mathbf{k}_{3}} (\hat{\eta}_{{\rm A}, \mathbf{k}_{1}} \hat{\eta}^{\dagger}_{\Phi, \mathbf{k}_{2}} \hat{\eta}^{\dagger}_{\Phi, \mathbf{k}_{3}} + \mathrm{h.c.}) + \hat{H}^{(3)}_{{\rm other}},
\end{align}
where $\hat{H}^{(3)}_{{\rm other}}$ collects the remaining cubic terms that do not contribute to the damping of the Higgs mode. Combining the diagonalized quadratic term and cubic interaction term and omitting higher-order terms as well as the constant vacuum-energy shift, we arrive at Eq.~\eqref{eq:diagHam}.
\end{widetext}

\section{Explicit coefficients in the effective Hamiltonian} \label{A2}
In this Appendix, we give the coefficients in each partial Hamiltonian $\hat{H}^{(i)}$ for $i = 1, 2, 3$.

The coefficient in the $\hat{H}^{(1)}$ is given by
\begin{align}
  A_1 = 4J z_{\mathrm{eff}} s_{1} c_{1} \sqrt{M} \left( c^{2}_{1} - s^{2}_{1} - u\right) = 0.
\end{align}
This implies $u=c^{2}_{1} - s^{2}_{1}$, where we recall that $u=\frac{\Delta}{4 J z_{\rm eff}}$.
The coefficients in the $\hat{H}^{(2)}$ are given by
\begin{align}
  B_{1} &=   4J z_{\mathrm{eff}}, \\
  B_{2} &= - J z_{\mathrm{eff}} u^{2} \gamma_{\mathbf{k}}, \\
  C_{1} &= 2J z_{\mathrm{eff}} (1 + u), \\
  C_{2} &= \frac{J z_{\mathrm{eff}}}{2} (1 + u) \gamma_{\mathbf{k}}.
\end{align}

The coefficients in the $\hat{H}^{(3)}$ are given by
\begin{align}
  D_1 &= - \frac{2}{\sqrt{M}} J z_{\mathrm{eff}} u\sqrt{1-u^2}\gamma_{\mathbf{k}_3} \delta_{\mathbf{k}_1 + \mathbf{k}_3, \mathbf{k}_2},\\
  D_2 &= - \frac{4}{\sqrt{M}} J z_{\mathrm{eff}} u\sqrt{1-u^2} \gamma_{\mathbf{k}_3} \delta_{\mathbf{k}_1 + \mathbf{k}_3, \mathbf{k}_2},\\
  D_3 &= \frac{1}{\sqrt{M}} J z_{\mathrm{eff}} \sqrt{1-u^2}  \gamma_{\mathbf{k}_3} \delta_{\mathbf{k}_1, \mathbf{k}_2 + \mathbf{k}_3}.
\end{align}
\begin{widetext}
The coefficient in the $\mathcal{S}^{(3)}_{{\rm damping}}$ is given by
\begin{align}
  \mathcal{M}_{\mathbf{k}_1, \mathbf{k}_2, \mathbf{k}_3} =  \frac{1}{\sqrt{M}} \delta_{\mathbf{k}_1, \mathbf{k}_2 + \mathbf{k}_3} \Bigg[ &- 2J z_{\mathrm{eff}} u \sqrt{1 - u^2} \gamma_{\mathbf{k}_1} (u_{\mathrm{A}, \mathbf{k}_1} + v_{\mathrm{A}, \mathbf{k}_1}) (u_{\Phi, \mathbf{k}_2} v_{\Phi, \mathbf{k}_3} + v_{\Phi, \mathbf{k}_2} u_{\Phi, \mathbf{k}_3}) \nonumber \\
  & + J z_{\mathrm{eff}} \sqrt{1 - u^2} \gamma_{\mathbf{k}_2}  (u_{\mathrm{A}, \mathbf{k}_1} u_{\Phi, \mathbf{k}_3} - v_{\mathrm{A}, \mathbf{k}_1} v_{\Phi, \mathbf{k}_3}) (u_{\Phi, \mathbf{k}_2} - v_{\Phi, \mathbf{k}_2}) \nonumber \\
  & + J z_{\mathrm{eff}} \sqrt{1 - u^2} \gamma_{\mathbf{k}_3}(u_{\mathrm{A}, \mathbf{k}_1} u_{\Phi, \mathbf{k}_2} - v_{\mathrm{A}, \mathbf{k}_1} v_{\Phi, \mathbf{k}_2}) (u_{\Phi, \mathbf{k}_3} - v_{\Phi, \mathbf{k}_3}) \Bigg].
\end{align}

\section{Supplement on the derivation of the approximate expression for the Beliaev damping rate} \label{App:Damping}
In this appendix, we present the intermediate steps leading from Eq.~\eqref{BeliaevDampingRate} to Eq.~\eqref{IntegratingDampingRate}. Near the quantum critical point, the on-shell decay of a zero-momentum Higgs mode is dominated by long-wavelength NG modes, because the energy conservation condition $\Delta_{\rm h}=2\mathcal{E}_{\Phi,\mathbf{k}}$ selects small momenta when $\Delta_{\rm h}$ is small. This justifies the low-energy approximation for both the NG dispersion and the Bogoliubov coefficients.
\begin{align}
  u_{\mathrm{A}, \mathbf{k}} \simeq v_{\mathrm{A}, \mathbf{k}} \simeq \frac{1}{\sqrt{2 \bar{\Delta}_{\mathrm{h}}}}, \quad 
  u_{\Phi, \mathbf{k}} \simeq -v_{\Phi, \mathbf{k}} \simeq \sqrt{\frac{u + 1}{4 \bar{g}_{\mathrm{ng}} |\mathbf{k}|^{\frac{\alpha - d}{2}}}}.
\end{align}
By substituting these expressions into Eq.~\eqref{BeliaevDampingRate}, we obtain
\begin{align}
  \Gamma_{{\rm A}, \mathbf{k} = \mathbf{0}} = (J z_{{\rm eff}})^{2} \frac{\pi (1 - u^{2}) (1 + u)^{4}}{4 \bar{\Delta}_{{\rm h}} \bar{g}_{{\rm ng}}^{2}} \int \frac{d^{d} \mathbf{k}}{(2\pi)^{d}} \frac{1}{|\mathbf{k}|^{\alpha - d}} \coth\left( \frac{\beta \mathcal{E}_{\Phi, \mathbf{k}}}{2} \right) \delta(\Delta_{\rm h} - 2\mathcal{E}_{\Phi, \mathbf{k}}), \label{App:Beliaev_v2_1}
\end{align}
where we have replaced the Bose distribution function with the hyperbolic cotangent form, $1 + 2 n_{\mathrm{B}}(\mathcal{E}) = \coth(\beta \mathcal{E} / 2)$.

In the long-wavelength limit ($|\mathbf{k}| \ll 1$), we employ the dispersion $\mathcal{E}_{\Phi, \mathbf{k}} \simeq g_{\rm ng} |\mathbf{k}|^{\frac{\alpha-d}{2}}$. Transforming the integral into hyperspherical coordinates yields
\begin{align}
  \Gamma_{{\rm A}, \mathbf{k} = \mathbf{0}} = (J z_{{\rm eff}})^{2} \frac{\pi (1 - u^{2}) (1 + u)^{4}}{4 \bar{\Delta}_{{\rm h}} \bar{g}_{{\rm ng}}^{2}} \frac{\Omega_d}{(2\pi)^d} \int_{0}^{\infty} d|\mathbf{k}| |\mathbf{k}|^{2d-\alpha-1} \coth\left( \frac{\beta g_{\rm ng} |\mathbf{k}|^{\frac{\alpha-d}{2}}}{2} \right) \frac{\delta\left( |\mathbf{k}|^{\frac{\alpha-d}{2}} - \frac{\Delta_{\rm h}}{2g_{\rm ng}} \right)}{2g_{\rm ng}},
\end{align}
where $\Omega_d = 2\pi^{d/2}/\Gamma(d/2)$ is the surface area of a unit sphere in $d$ dimensions. 
To evaluate the integral, we introduce the change of variables $s = |\mathbf{k}|^{\frac{\alpha-d}{2}}$, which gives $d|\mathbf{k}| = \frac{2}{\alpha-d} s^{\frac{2}{\alpha-d}-1} ds$. The integral is then rewritten as
\begin{align}
  \Gamma_{{\rm A}, \mathbf{k} = \mathbf{0}} = (J z_{{\rm eff}})^{2} \frac{\pi (1 - u^{2}) (1 + u)^{4}}{8 \bar{\Delta}_{{\rm h}} \bar{g}_{{\rm ng}}^{2} g_{\rm ng}} \frac{\Omega_d}{(2\pi)^d} \int_{0}^{\infty} ds \left( \frac{2}{\alpha-d} \right) s^{\frac{2(2d-\alpha)}{\alpha-d} - 1} \coth\left( \frac{\beta g_{{\rm ng}} s}{2} \right) \delta\left( s - \frac{\Delta_{\rm h}}{2g_{\rm ng}} \right).
\end{align}
Performing the integration over $s$ with the aid of the delta function and using the on-shell condition $s=\frac{\Delta_{\rm h}}{2g_{\rm ng}}$, we obtain
\begin{align}
  \Gamma_{{\rm A}, \mathbf{k} = \mathbf{0}}
  &\simeq \frac{\pi J z_{\mathrm{eff}} }{8 (\alpha - d)} \frac{\Omega_d \sqrt{1 - u^2} (1 + u)}{(2\pi)^d A^{3/2}} \left(\frac{\Delta_{\mathrm{h}}}{2 g_{\mathrm{ng}}}\right) ^{\frac{5d - 3 \alpha}{\alpha - d}} \coth \left( \frac{\beta \Delta_{\mathrm{h}}}{4} \right),
\end{align}
which reproduces Eq.~\eqref{IntegratingDampingRate}.
\end{widetext}

\bibliographystyle{apsrev4-2}

\begin{thebibliography}{99}
\bibitem{Defenu2023} N. Defenu, T. Donner, T. Macri, G. Pagano, S. Ruffo, and A. Trombettoni, Long-range interacting quantum systems,  \href{https://doi.org/10.1103/RevModPhys.95.035002}{Rev. Mod. Phys. {\bf 95}, 035002 (2023)}.
  \bibitem{Browaeys2020} A. Browaeys and T. Lahaye, Many-body physics with individually controlled Rydberg atoms, \href{https://doi.org/10.1038/s41567-019-0733-z}{Nat. Phys. \textbf{16}, 132 (2020)}.
  \bibitem{Morgado2021} M. Morgado, and S. Whitlock, Quantum simulation and computing with Rydberg-interacting qubits, \href{https://doi.org/10.1116/5.0036562}{AVS Quantum Sci. \textbf{3}, 023501 (2021)}.
  \bibitem{Lahaye2009} T. Lahaye, C. Menotti, L. Santos, M. Lewenstein, and T. Pfau, The physics of dipolar bosonic quantum gases, \href{https://doi.org/10.1088/0034-4885/72/12/126401}{Rep. Prog. Phys. \textbf{72}, 126401 (2009)}.
  \bibitem{Chomaz2022} L. Chomaz, I. Ferrier-Barbut, F. Ferlaino, B. Laburthe-Tolra, B. L. Lev, and T. Pfau, Dipolar physics: a review of experiments with magnetic quantum gases, \href{https://doi.org/10.1088/1361-6633/aca814}{Rep. Prog. Phys. {\bf 86}, 026401 (2022)}.
  \bibitem{Blatt2012} R. Blatt and C. F. Roos, Quantum simulations with trapped ions, \href{https://doi.org/10.1038/nphys2252}{Nat. Phys. {\bf 8}, 277 (2012)}.
  \bibitem{Monroe2021} C. Monroe, W. C. Campbell, L.-M. Duan, Z.-X. Gong, A. V. Gorshkov, P. W. Hess, R. Islam, K. Kim, N. M. Linke, G. Pagano, P. Richerme, C. Senko, and N. Y. Yao, Programmable quantum simulations of spin systems with trapped ions, \href{https://doi.org/10.1103/RevModPhys.93.025001}{Rev. Mod. Phys. \textbf{93}, 025001 (2021)}.
  \bibitem{Labuhn2016} H. Labuhn, D. Barredo, S. Ravets, S. de L\'{e}s\'{e}leuc, T. Macr\`{i}, T. Lahaye, and A. Browaeys, Tunable two-dimensional arrays of single Rydberg atoms for realizing quantum Ising models, \href{https://doi.org/10.1038/nature18274}{Nature (London) \textbf{534}, 667-670 (2016)}.
  \bibitem{Bernien2017} H. Bernien, S. Schwartz, A. Keesling, H. Levine, A. Omran, H. Pichler, S. Choi, A. S. Zibrov, M. Endres, M. Greiner, V. Vuleti\'{c}, and M. D. Lukin, Probing many-body dynamics on a 51-atom quantum simulator, \href{https://doi.org/10.1038/nature24622}{Nature (London) \textbf{551}, 579-584 (2017)}.
  \bibitem{Scholl2021} P. Scholl, M. Schuler, H. J. Williams, A. A. Eberharter, D. Barredo, K. N. Schymik, V. Lienhard, L. P. Henry, T. C. Lang, T. Lahaye, A. M. L\"{a}uchli, and A. Browaeys, Quantum simulation of 2D antiferromagnets with hundreds of Rydberg atoms, \href{https://doi.org/10.1038/s41586-021-03585-1}{Nature (London) \textbf{595} , 233-238 (2021)}.
  \bibitem{Sylvain2019} S. de L\'{e}s\'{e}leuc, V. Lienhard, P. Scholl, D. Barredo, S. Weber, N. Lang, H. P. B\"{u}chler, T. Lahaye, A. Browaeys, Observation of a symmetry-protected topological phase of interacting bosons with Rydberg atoms, \href{https://doi.org/10.1126/science.aav9105}{Science \textbf{365}, 775-780 (2019)}.
  \bibitem{Chen2023} C. Chen, G. Bornet, M. Bintz, G. Emperauger, L. Leclerc, V. S. Liu, P. Scholl, D. Barredo, J. Hauschild, S. Chatterjee, M. Schuler, A. M. L\"{a}uchli, M. P. Zaletel, T. Lahaye, N. Y. Yao, and A. Browaeys, Continuous symmetry breaking in a two-dimensional Rydberg array, \href{https://doi.org/10.1038/s41586-023-05859-2}{Nature (London) \textbf{616}, 691-695 (2023)}.
  \bibitem{Chen2025} C. Chen, G. Emperauger, G. Bornet, F. Caleca, B. G\'{e}ly, M. Bintz, S. Catterjee, V. Liu, D. Barredo, N. Y. Yao, T. Lahaye, F. Mezzacapo, T. Roscilde, and A. Browaeys, Spectroscopy of elementary excitations from quench dynamics in a dipolar XY Rydberg simulator, \href{https://doi.org/10.1126/science.adn0618}{Science \textbf{389}, 483-487 (2025)}.
  \bibitem{Chew2022} Y. Chew, T. Tomita, T. P. Mahesh, S. Sugawa, S. de L\'{e}s\'{e}leuc, and K. Ohmori, Ultrafast energy exchange between two single Rydberg atoms on a nanosecond timescale, \href{https://doi.org/10.1038/s41566-022-01047-2}{Nat. Photonics \textbf{16}, 724 (2022)}.
  \bibitem{Qiao2025} M. Qiao, G. Emperauger, C. Chen, L. Homeier, S. Hollerith, G. Bornet, R. Martin, B. G\'{e}ly, L. Klein, D. Barredo, S. Geier, N.-C. Chiu, F. Grusdt, A. Bohrdt, T. Lahaye, and  A. Browaeys, Realization of a doped quantum antiferromagnet in a Rydberg tweezer array, \href{https://doi.org/10.1038/s41586-025-09377-1}{Nature (London) \textbf{644}, 889-895 (2025)}. 
  \bibitem{Yoshida2024} T. Yoshida, M. Kunimi, and T. Nikuni, Proposal for experimental realization of quantum spin chains with quasiperiodic interaction using Rydberg atoms, \href{https://arxiv.org/abs/2409.08497}{arXiv:2409.08497}.
  \bibitem{Moegerle2025} J. M\"{o}gerle, K. Brechtelsbauer, A.T. G.-Caballero, J. Prior, G. Emperauger, G. Bornet, C. Chen, T. Lahaye, A. Browaeys, and H. P. B\"{u}chler, Spin-1 Haldane Phase in a Chain of Rydberg Atoms, \href{https://link.aps.org/doi/10.1103/PRXQuantum.6.020332}{PRX Quantum {\bf 6}, 020332 (2025)}.
  \bibitem{Pekker2015} D. Pekker and C. M. Varma, Amplitude/Higgs Modes in Condensed Matter Physics, \href{https://www.annualreviews.org/content/journals/10.1146/annurev-conmatphys-031214-014350}{Annu. Rev. Condens. Matter Phys. \textbf{6}, 269 (2015)}.
  \bibitem{Shimano2020} R. Shimano and N. Tsuji, Higgs Mode in Superconductors, \href{https://doi.org/10.1146/annurev-conmatphys-031119-050813}{Annu. Rev. Condens. Matter Phys. \textbf{11}, 103-124 (2020)}.
  \bibitem{Tsuji2024} N. Tsuji, I. Danshita, and S. Tsuchiya, Higgs and Nambu-Goldstone modes in condensed matter physics, \href{https://doi.org/10.1016/B978-0-323-90800-9.00256-0}{Encycl. Condens. Matter Phys. \textbf{1}, 174 (2024)}.
  \bibitem{Sooryakumar1980} R. Sooryakumar and M. V. Klein, Raman Scattering by Superconducting-Gap Excitations and Their Coupling to Charge-Density Waves, \href{https://doi.org/10.1103/PhysRevLett.45.660}{Phys. Rev. Lett. {\bf 45}, 660 (1980)}.
  \bibitem{Sooryakumar1981} R. Sooryakumar and M. V. Klein, Raman scattering from superconducting gap excitations in the presence of a magnetic field, \href{https://doi.org/10.1103/PhysRevB.23.3213}{Phys. Rev. B {\bf 23}, 3213 (1981)}.
  \bibitem{Balseiro1980} C. A. Balseiro and L. M. Falicov, Phonon Raman Scattering in Superconductors, \href{https://doi.org/10.1103/PhysRevLett.45.662}{Phys. Rev. Lett. {\bf 45}, 662 (1980)}.
  \bibitem{Littlewood1981} P. B. Littlewood and C. M. Varma, Gauge-Invariant Theory of the Dynamical Interaction of Charge Density Waves and Superconductivity, \href{https://doi.org/10.1103/PhysRevLett.47.811}{Phys. Rev. Lett. {\bf 47}, 811 (1981)}.
  \bibitem{Littlewood1982} P. B. Littlewood and C. M. Varma, Amplitude collective modes in superconductors and their coupling to charge-density waves, \href{https://doi.org/10.1103/PhysRevB.26.4883}{Phys. Rev. B {\bf 26}, 4883 (1982)}.
  \bibitem{Measson2014} M.-A. M\'{e}asson, Y. Gallais, M. Cazayous, B. Clair, P. Rodi\`{e}re, L. Cario, and A. Sacuto, Amplitude Higgs mode in the $2H\ensuremath{-}{\text{NbSe}}_{2}$ superconductor, \href{https://doi.org/10.1103/PhysRevB.89.060503}{Phys. Rev. B {\bf 89}, 060503(R) (2014)}.
  \bibitem{Matsunaga2013} R. Matsunaga, Y. I. Hamada, K. Makise, Y. Uzawa, H. Terai, Z. Wang, and R. Shimano, Higgs Amplitude Mode in the BCS Superconductors Nb$_{1-x}$Ti$_x$N Induced by Terahertz Pulse Excitation, \href{https://doi.org/10.1103/PhysRevLett.111.057002}{Phys. Rev. Lett. {\bf 111}, 057002 (2013)}.
  \bibitem{Matsunaga2014} R. Matsunaga, N. Tsuji, H. Fujita, A. Sugioka, K. Makise, Y. Uzawa, H. Terai, Z. Wang, H. Aoki, and R. Shimano, Light-induced collective pseudospin precession resonating with Higgs mode in a superconductor, \href{https://doi.org/10.1126/science.1254697}{Science {\bf 345}, 6201 (2014)}.
  \bibitem{Sherman2015} D. Sherman, U. S. Pracht, B. Gorshunov, S. Poran, J. Jesudasan, M. Chand, P. Raychaudhuri, M. Swanson, N. Trivedi, A. Auerbach, M. Scheffler, A. Frydman, and M. Dressel, The Higgs mode in disordered superconductors close to a quantum phase transition, \href{https://doi.org/10.1038/nphys3227}{Nat. Phys. {\bf 11}, 188 (2015)}.
  \bibitem{Matsunaga2017} R. Matsunaga, N. Tsuji, K. Makise, H. Terai, H. Aoki, and R. Shimano, Polarization-resolved terahertz third-harmonic generation in a single-crystal superconductor NbN: Dominance of the Higgs mode beyond the BCS approximation, \href{https://doi.org/10.1103/PhysRevB.96.020505}{Phys. Rev. B {\bf 96}, 020505(R) (2017)}.
  \bibitem{Katsumi2024} K. Katsumi, J. Fiore, M. Udina, R. Romero, D. Barbalas, J. Jesudasan, P. Raychaudhuri, G. Seibold, L. Benfatto, and N. P. Armitage, Revealing Novel Aspects of Light-Matter Coupling by Terahertz Two-Dimensional Coherent Spectroscopy: The Case of the Amplitude Mode in Superconductors, \href{https://doi.org/10.1103/PhysRevLett.132.256903}{Phys. Rev. Lett. {\bf 132}, 256903 (2024)}.
  \bibitem{Katsumi2025} K. Katsumi, J. Liang, R. Romero, K. Chen, X. Xi, and N. P. Armitage, Amplitude Mode in a Multigap Superconductor MgB$_2$ Investigated by Terahertz Two-Dimensional Coherent Spectroscopy, \href{https://doi.org/10.1103/g5rp-6vb1}{Phys. Rev. Lett. {\bf 135}, 036902 (2025)}.
  \bibitem{Katsumi2018} K. Katsumi, N. Tsuji, Y. I. Hamada, R. Matsunaga, J. Schneeloch, R. D. Zhong, G. D. Gu, H. Aoki, Y. Gallais, and R. Shimano, Higgs Mode in the $d$-Wave Superconductor Bi$_2$Sr$_2$CaCu$_2$O$_{8+x}$ Driven by an Intense Terahertz Pulse, \href{https://doi.org/10.1103/PhysRevLett.120.117001}{Phys. Rev. Lett. {\bf 120}, 117001 (2018)}.
  \bibitem{Chu2020} H. Chu, M.-J. Kim, K. Katsumi, S. Kovalev, R. D. Dawson, L. Schwarz, N. Yoshikawa, G. Kim, D. Putzky, Z. Z. Li, H. Raffy, S. Germanskiy, J.-C. Deinert, N. Awari, I. Ilyakov, B. Green, M. Chen, M. Bawatna, G. Cristiani, G. Logvenov, Y. Gallais, A. V. Boris, B. Keimer, A. P. Schnyder, D. Manske, M. Gensch, Z. Wang, R. Shimano, and S. Kaiser, Phase-resolved Higgs response in superconducting cuprates, \href{https://doi.org/10.1038/s41467-020-15613-1}{Nat. Commun. {\bf 11}, 1793 (2020)}.
  \bibitem{Ruegg2008} Ch. R\"uegg, B. Normand, M. Matsumoto, A. Furrer, D. F. McMorrow, K. W. Kramer, H. U. Gudel, S. N. Gvasaliya, H. Mutka, and M. Boehm, Quantum Magnets under Pressure: Controlling Elementary Excitations in TlCuCl$_3$, \href{https://doi.org/10.1103/PhysRevLett.100.205701}{Phys. Rev. Lett. {\bf 100}, 205701 (2008)}.
  \bibitem{Merchant2014} P. Merchant, B. Normand, K. W. Kr\"amer, M. Boehm, D. F. McMorrow, and Ch. R\"uegg, Quantum and classical criticality in a dimerized quantum antiferromagnet, \href{https://doi.org/10.1038/nphys2902}{Nat. Phys. {\bf 10}, 373 (2014)}.
  \bibitem{Kuroe2012} H. Kuroe, N. Takami, N. Niwa, T. Sekine, M. Matsumoto, F. Yamada, H. Tanaka, and K. Takemura, Longitudinal magnetic excitation in KCuCl$_3$ studied by Raman scattering under hydrostatic pressures, \href{http://doi.org/10.1088/1742-6596/400/3/032042}{J. Phys.: Conf. Series {\bf 400}, 032042 (2012)}.
  \bibitem{Demsar1999} J. Demsar, K. Biljakovi\'{c}, and D. Mihailovic, Single Particle and Collective Excitations in the One-Dimensional Charge Density Wave Solid K$_{0.3}$MoO$_3$ Probed in Real Time by Femtosecond Spectroscopy, \href{https://doi.org/10.1103/PhysRevLett.83.800}{Phys. Rev. Lett. {\bf 83}, 800 (1999)}.
  \bibitem{Schaefer2014} H. Schaefer, V. V. Kabanov, and J. Demsar, Collective modes in quasi-one-dimensional charge-density wave systems probed by femtosecond time-resolved optical studies, \href{https://doi.org/10.1103/PhysRevB.89.045106}{Phys. Rev. B {\bf 89}, 045106 (2014)}.
  \bibitem{Yusupov2010} R. Yusupov, T. Mertelj, V. V. Kabanov, S. Brazovskii, P. Kusar, J.-H. Chu, I. R. Fisher, and D. Mihailovic, Coherent dynamics of macroscopic electronic order through a symmetry breaking transition, \href{https://doi.org/10.1038/nphys1738}{Nat. Phys. {\bf 6}, 681 (2010)}.
  \bibitem{Mertelj2013} T. Mertelj, P. Kusar, V. V. Kabanov, P. Giraldo-Gallo, I. R. Fisher, and D. Mihailovic, Incoherent Topological Defect Recombination Dynamics in TbTe$_3$, \href{https://doi.org/10.1103/PhysRevLett.110.156401}{Phys. Rev. Lett. {\bf 110}, 156401 (2013)}.
  \bibitem{Avenel1980} O. Avenel, E. Varoquaux, and H. Ebisawa, Field Splitting of the New Sound Attenuation Peak in $^3$He-$B$, \href{https://doi.org/10.1103/PhysRevLett.45.1952}{Phys. Rev. Lett. {\bf 45}, 1952 (1980)}.
  \bibitem{Collett2013} C. A. Collett, J. Pollanen, J. I. A. Li, W. J. Gannon, and W. P. Halperin, Zeeman Splitting and Nonlinear Field-Dependence in Superfluid $^3$He, \href{https://doi.org/10.1007/s10909-012-0692-6}{J. Low Temp. Phys. {\bf 171}, 214 (2013)}.
  \bibitem{Bissbort2011} U. Bissbort, S. G\"{o}tze, Y. Li, J. Heinze, J. S. Krauser, M. Weinberg, C. Becker, K. Sengstock, and W. Hofstetter, Detecting the Amplitude Mode of Strongly Interacting Lattice Bosons by Bragg Scattering, \href{https://doi.org/10.1103/PhysRevLett.106.205303}{Phys. Rev. Lett. {\bf 106}, 205303 (2011)}.
  \bibitem{Endres2012} M. Endres, T. Fukuhara, D. Pekker, M. Cheneau, P. Schau{\ss}, C. Gross, E. Demler, S. Kuhr, and I. Bloch, The `Higgs’ amplitude mode at the two-dimensional superfluid/Mott insulator transition, \href{https://doi.org/10.1038/nature11255}{Nature (London) \textbf{487}, 454 (2012)}.
  \bibitem{Leonard2017} J. L\'eonard, A. Morales, P. Zupancic, T. Donner, and T. Esslinger, Monitoring and manipulating Higgs and Goldstone modes in a supersolid quantum gas, \href{https://doi.org/10.1126/science.aan2608}{Science {\bf 358}, 1415 (2017)}.
  \bibitem{Hertkorn2019} J. Hertkorn, F. B\"ottcher, M. Guo, J. N. Schmidt, T. Langen, H. P. B\"uchler, and T. Pfau, Fate of the Amplitude Mode in a Trapped Dipolar Supersolid, \href{https://doi.org/10.1103/PhysRevLett.123.193002}{Phys. Rev. Lett. {\bf 123}, 193002 (2019)}.
  \bibitem{Behrle2018} A. Behrle, T. Harrison, J. Kombe, K. Gao, M. Link, J. S. Bernier, C. Kollath, and M. K\"ohl, Higgs mode in a strongly interacting fermionic superfluid, \href{ https://doi.org/10.1038/s41567-018-0128-6}{Nat. Phys. {\bf 14}, 781 (2018)}.
  \bibitem{Dyke2024} P. Dyke, S. Musolino, H. Kurkjian, D. J. M. Ahmed-Braun, A. Pennings, I. Herrera, S. Hoinka, S. J. J. M. F. Kokkelmans, V. E. Colussi, and C. J. Vale, Higgs Oscillations in a Unitary Fermi Superfluid, \href{https://doi.org/10.1103/PhysRevLett.132.223402}{Phys. Rev. Lett. {\bf 132}, 223402 (2024)}. 
  \bibitem{Kell2024} A. Kell, M. Breyer, D. Eberz, and M. K\"ohl, Exciting the Higgs Mode in a Strongly Interacting Fermi Gas by Interaction Modulation, \href{https://doi.org/10.1103/PhysRevLett.133.150403}{Phys. Rev. Lett. {\bf 133}, 150403 (2024)}.
  \bibitem{Cabrera2025} C. R. Cabrera, R. Henke, L. Broers, J. Skulte, H. P. Ojeda Collado, H. Biss, L. Mathey, and H. Moritz, Hybridization of the amplitude mode in a confined fermionic superfluid, \href{https://doi.org/10.1103/5cv2-f4ng}{Phys. Rev. Research {\bf 7}, L032018 (2025)}.
  \bibitem{Higgs1964} P. W. Higgs, Broken Symmetries and the Masses of Gauge Bosons, \href{https://link.aps.org/doi/10.1103/PhysRevLett.13.508}{Phys. Rev. Lett. \textbf{13}, 508 (1964)}.
  \bibitem{Altman2002} E. Altman and A. Auerbach, Oscillating Superfluidity of Bosons in Optical Lattices, \href{https://link.aps.org/doi/10.1103/PhysRevLett.89.250404}{Phys. Rev. Lett. \textbf{89}, 250404 (2002)}.
  \bibitem{Podolsky2011} D. Podolsky, A. Auerbach, and D. P. Arovas, Visibility of the amplitude (Higgs) mode in condensed matter, \href{https://doi.org/10.1103/PhysRevB.84.174522}{Phys. Rev. B {\bf 84}, 174522 (2011)}.
  \bibitem{Pollet2012} L. Pollet and N. Prokof'ev, Higgs Mode in a Two-Dimensional Superfluid, \href{https://doi.org/10.1103/PhysRevLett.109.010401}{Phys. Rev. Lett. {\bf 109}, 010401 (2012)}.
  \bibitem{Podolsky2012} D. Podolsky and S. Sachdev, Spectral functions of the Higgs mode near two-dimensional quantum critical points, \href{https://doi.org/10.1103/PhysRevB.86.054508}{Phys. Rev. B {\bf 86}, 054508 (2012)}.
  \bibitem{Rancon2014} A. Ran\c{c}on and N. Dupuis, Higgs amplitude mode in the vicinity of a $(2+1)$-dimensional quantum critical point, \href{https://doi.org/10.1103/PhysRevB.89.180501}{Phys. Rev. B {\bf 89}, 180501(R) (2014)}.
  \bibitem{Bruno2001} P. Bruno, Absence of Spontaneous Magnetic Order at Nonzero Temperature in One- and Two-Dimensional Heisenberg and $XY$ Systems with Long-Range Interactions, \href{https://doi.org/10.1103/PhysRevLett.87.137203}{Phys. Rev. Lett. {\bf 87}, 137203 (2001)}.
  \bibitem{Sbierski2024} B. Sbierski, M. Bintz, S. Chatterjee, M. Schuler, N. Y. Yao, and L. Pollet, Magnetism in the two-dimensional dipolar XY model, \href{https://doi.org/10.1103/PhysRevB.109.144411}{Phys. Rev. B {\bf 109}, 144411 (2024)}.
  \bibitem{Nagao2016} K. Nagao and I. Danshita, Damping of the Higgs and Nambu-Goldstone modes of superfluid Bose gases at finite temperatures, \href{https://doi.org/10.1093/ptep/ptw061}{Prog. Theor. Exp. Phys. \textbf{2016}, 063I01 (2016)}.
  \bibitem{Nagao2016b} K. Nagao, Analyses of damping of collective excitations in strongly correlated Bose gases in optical lattices on the basis of the $XY$ model with single-ion anisotropy, \href{https://doi.org/10.14989/225160}{Bussei Kenkyu {\bf 5}, 053602 (2016)} [in Japanese].
  \bibitem{Ravets2014} S. Ravets, H. Labuhn, D. Barredo, L. B\'eguin, T. Lahaye, and A. Browaeys, Coherent dipole-dipole coupling between two single Rydberg atoms at an electrically-tuned F\"{o}rster resonance, \href{https://doi.org/10.1038/nphys3119}{Nat. Phys. {\bf 10}, 914 (2014)}.
  \bibitem{Danshita2011} I. Danshita and A. Polkovnikov, Superfluid-to-Mott-insulator transition in the one-dimensional Bose-Hubbard model for arbitrary integer filling factors, \href{https://doi.org/10.1103/PhysRevA.84.063637}{Phys. Rev. A {\bf 84}, 063637 (2011)}.
  \bibitem{Teichmann2009} N. Teichmann, D. Hinrichs, M. Holthaus, and A. Eckardt, Bose-Hubbard phase diagram with arbitrary integer filling, \href{https://doi.org/10.1103/PhysRevB.79.100503}{Phys. Rev. B {\bf 79}, 100503(R) (2009)}.
  \bibitem{Teichmann2009b} N. Teichmann, D. Hinrichs, M. Holthaus, and A. Eckardt, Process-chain approach to the Bose-Hubbard model: Ground-state properties and phase diagram, \href{https://doi.org/10.1103/PhysRevB.79.224515}{Phys. Rev. B {\bf 79}, 224515 (2009)}.  
  \bibitem{Danshita2010} I. Danshita and D. Yamamoto, Critical velocity of flowing supersolids of dipolar Bose gases in optical lattices, \href{https://doi.org/10.1103/PhysRevA.82.013645}{Phys. Rev. A \textbf{82}, 013645 (2010)}.
  \bibitem{Huber2007} S. D. Huber, E. Altman, H. P. B\"{u}chler, and G. Blatter, Dynamical properties of ultracold bosons in an optical lattice, \href{https://doi.org/10.1103/PhysRevB.75.085106}{Phys. Rev. B \textbf{75}, 085106 (2007)}.
  \bibitem{Wang2021} J. Wang, Y. Deng, and W. Zhang, Topological Higgs amplitude modes in strongly interacting superfluids, \href{https://doi.org/10.1103/PhysRevA.104.043328}{Phys. Rev. A {\bf 104}, 043328 (2021)}.
  \bibitem{Holstein1940} T. Holstein and H. Primakoff, Field Dependence of the Intrinsic Domain Magnetization of a Ferromagnet, \href{https://link.aps.org/doi/10.1103/PhysRev.58.1098}{Phys. Rev. \textbf{58}, 1098 (1940)}.
  \bibitem{Abrikosov1975} A. A. Abrikosov, L. P. Gorkov, and I. E. Dzyaloshinski, \textit{Methods of Quantum Field Theory in Statistical Physics} (Dover, New York, 1975).
  \bibitem{Lifshitz1980} E. M. Lifshitz and L. P. Pitaevskii, \textit{Statistical Physics}, Part 2 (Pergamon, Oxford, 1980).
  \bibitem{Altland2010} A. Altland and B. D. Simons, \textit{Condensed Matter Field Theory}, 2nd ed. (Cambridge University Press, Cambridge, 2010).
  \bibitem{Diessel2023} O. K. Diessel, S. Diehl, N. Defenu, A. Rosch, and A. Chiocchetta, Generalized Higgs mechanism in long-range-interacting quantum systems, \href{https://doi.org/10.1103/PhysRevResearch.5.033038}{Phys. Rev. Research \textbf{5}, 033038 (2023)}.
  \bibitem{Song2023} M. Song, J. Zhao, C. Zhou, and Z.-Y. Meng, Dynamical properties of quantum many-body systems with long-range interactions, \href{https://doi.org/10.1103/PhysRevResearch.5.033046}{Phys. Rev. Research {\bf 5}, 033046 (2023)}.
  \bibitem{Adelhardt2025} P. Adelhardt, A. Duft, and K. P. Schmidt, Quantum-critical and dynamical properties of the XXZ bilayer with long-range interactions, \href{https://doi.org/10.1103/PhysRevB.111.024409}{Phys. Rev. B {\bf 111}, 024409 (2025)}.
  \bibitem{Yabuuchi2025} Y. Yabuuchi and I. Danshita, Stability of current-carrying states in hard-core bosons with long-range hopping on a square lattice, \href{https://doi.org/10.48550/arXiv.2511.14260}{arXiv:2511.14260} [\href{https://doi.org/10.1103/xvx2-n5j2}{Phys. Rev. A} (to be published)].
  \bibitem{Peter2012} D. Peter, S. M\"{u}ller, S. Wessel, and H. P. B\"{u}chler, Anomalous Behavior of Spin Systems with Dipolar Interactions, \href{https://doi.org/10.1103/PhysRevLett.109.025303}{Phys. Rev. Lett. \textbf{109}, 025303 (2012)}.
  \bibitem{Gao2021} H. Gao, F. Schlawin, and D. Jaksch, Higgs mode stabilization by photoinduced long-range interactions in a superconductor, \href{https://doi.org/10.1103/PhysRevB.104.L140503}{Phys. Rev. B {\bf 104}, L140503 (2021)}.
  \bibitem{Manovitz2025} T. Manovitz, S. H. Li, S. Ebadi, R. Samajdar, A. A. Geim, S. J. Evered, D. Bluvstein, H. Zhou, N. U. Koyluoglu, J. Feldmeier, P. E. Dolgirev, N. Maskara, M. Kalinowski, S. Sachdev, D. A. Huse, M. Greiner, V. Vuleti\'c, and M. D. Lukin, Quantum coarsening and collective dynamics on a programmable simulator, \href{https://doi.org/10.1038/s41586-024-08353-5}{Nature (London) {\bf 638}, 86 (2025)}.
  \bibitem{Barankov2004} R. A. Barankov, L. S. Levitov, and B. Z. Spivak, Collective Rabi Oscillations and Solitons in a Time-Dependent BCS Pairing Problem, \href{https://doi.org/10.1103/PhysRevLett.93.160401}{Phys. Rev. Lett. {\bf 93}, 160401 (2004)}.
  \bibitem{Yuzbashyan2006} E. A. Yuzbashyan, O. Tsyplyatyev, and B. L. Altshuler, Relaxation and Persistent Oscillations of the Order Parameter in Fermionic Condensates, \href{https://doi.org/10.1103/PhysRevLett.96.097005}{Phys. Rev. Lett. {\bf 96}, 097005 (2006)}.
  \bibitem{Tokimoto2019} J. Tokimoto, S. Tsuchiya, and T. Nikumi, Excitation of Higgs Mode in Superfluid Fermi Gas in BCS-BEC Crossover \href{https://doi.org/10.7566/JPSJ.88.023601}{J. Phys. Soc. Jpn. {\bf 88}, 023601 (2019)}.
  \bibitem{Collado2023} H. P. Ojeda Collado, N. Defenu, and J. Lorenzana, Engineering Higgs dynamics by spectral singularities, \href{https://doi.org/10.1103/PhysRevResearch.5.023011}{Phys. Rev. Research {\bf 5}, 023011 (2023)}.
  \bibitem{Nakayama2015} T. Nakayama, I. Danshita, T. Nikuni, and S. Tsuchiya, Fano resonance through Higgs bound states in tunneling of Nambu-Goldstone modes, \href{https://doi.org/10.1103/PhysRevA.92.043610}{Phys. Rev. A {\bf 92}, 043610 (2015)}.
  \bibitem{Danshita2017} I. Danshita and S. Tsuchiya, Localized Higgs modes of superfluid Bose gases in optical lattices: A Gutzwiller mean-field study, \href{https://doi.org/10.1103/PhysRevA.96.043606}{Phys. Rev. A {\bf 96}, 043606 (2017)}.
\end{thebibliography}

\end{document}